\documentclass[12pt]{iopart}

\usepackage{longtable}
\usepackage{iopams}  
\usepackage[usenames]{color}
\usepackage{comment}
\usepackage{ascmac}
\usepackage{graphicx}
\usepackage{caption}
\newcommand{\ket}[1]{\big\vert\, #1\, \big\rangle}
\newcommand{\bra}[1]{\big\langle\, #1\, \big\vert}

\begin{document}

\title[Generation of Twisted Gamma-Rays via Two-Photon Transitions]
{Generation of Twisted Gamma-Rays via Two-Photon Transition}

\author{Motomichi Tashiro$^1$, Noboru Sasao$^2$ and Minoru Tanaka$^3$}

\address{$^1$ Department of Applied Chemistry, Toyo University, %Kujirai 2100,
           Kawagoe, Saitama 350-8585, Japan}
\address{$^2$ Research Institute for Interdisciplinary Science,
           Okayama University, Okayama 700-8530, Japan}
\address{$^3$ Graduate School of Science, Osaka University, Toyonaka, Osaka 560-0043, Japan}
\ead{sasao@okayama-u.ac.jp}
\vspace{10pt}
\begin{indented}
\item[]March 2022
\end{indented}

\begin{abstract}
We present a new and efficient method of generating 
twisted gamma rays utilizing highly accelerated helium-like ions; 
they are excited to a chosen state by irradiating 
two optical lasers and emit a photon with orbital angular momentum in the deexcitation
process. 
We study its emission rate together with other properties such as background and photo-ionization processes.

\end{abstract}

%\maketitle

%--------------Sec 1--------------------%
\section{Introduction}
%--------------Sec 1--------------------%
% what ``twisted'' means, history
Light or, its quantum, photon is characterized by the linear momentum
(or wave vector) and the helicity in the case of plane wave.
It is also possible to describe the photon in terms of the multipole field
specified by the energy, the total angular momentum $J_\gamma$ and its projection 
$M_\gamma$ onto a quantization axis \cite{LLQED}.
Then, one may say that a photon has an orbital angular momentum (OAM)
as well as the spin angular momentum if $J_\gamma\ge 2$.

In the form of propagating beam, light with OAM is realized by 
the Laguerre-Gaussian beam, which solves the wave equation in 
the paraxial approximation, as Allen et al. discovered in 1992
\cite{Allen1992}. 
The wave front of such a beam is a helicoid, exhibiting a phase singularity and
a complete dark spot along the beam axis, thus mentioned as ``twisted'' or 
``vortex'' in the literature.
The OAM degree of freedom of twisted photons has provided a fresh view of
the light-matter interaction and novel applications
\cite{Shen2019,Padgett2017,Molina-Terriza2007,Torres-Torner2011}.
The range of applications is wide, including fundamental interaction between 
atoms and photons \cite{Babiker2019,GORGONE2019}, 
quantum optics \cite{Mair2001,Leach2009}, 
micro manipulation of particles/materials \cite{He1995,Garces2003}, 
microscopy and imaging \cite{Swartzlander2001,Swartzlander2008,Furhapter2005}, 
optical data transmission \cite{Erhard2018,Gibson2004,Krenn2016}, 
and astrophysics \cite{Harwit2003,Tamburini2020,Maruyama2021}.

% realization in visible/optical region
Fork holograms, spiral phase plates \cite{X-Wang2018}, lens-based mode 
converters \cite{Beijersbergen1993} and q-plates \cite{Marrucci2006} are
popular methods employed to generate optical beams with OAM.
% X ray region
As for the X-ray region, high-harmonic radiation from a helical undulator 
\cite{Sasaki2008,Bahrdt2013,Kaneyasu2018}, and coherent emission from 
spirally-bunched electrons produced by combination of a laser and undulator 
\cite{Hemsing2012,Hemsing2013,Ribic2017} are apparently promising.
% gamma ray region
Proposed methods to generate twisted gamma rays fall into two classes
depending on the method of energy up-conversion:
Backward Compton scattering \cite{Jentschura2011PRD,Jentschura2011EPC,
Ivanov-Sebo2011,Stock2015,Petrillo2016,Taira2017,Chen2019}, 
and photon absorption-emission by accelerated partially stripped ions (PSIs)
\cite{Budker2020,Tanaka2021,Serbo2021}. 
The proposed gamma factory \cite{Bessonov2013,Krasny2019} is supposed for 
the latter scenario.
Compared with the backward Compton scattering, the latter process has an 
advantage of having much bigger fundamental cross section; 
the Rayleigh scattering cross section proportional to square of the resonant
wavelength versus the Thomson scattering cross section proportional to 
square of the classical electron radius.
% envisaged applications
Beams of twisted gamma rays are expected to be useful in the studies of 
nuclear structure, spin puzzle of nucleons \cite{Ivanov2011}, and astrophysical
phenomena associated with rotation and/or strong magnetic field. 

% H-like ion
In ref.~\cite{Tanaka2021}, a method to generate twisted photons using
hydrogen-like (H-like) ions is studied.
A laser beam of twisted photons in the visible region is irradiated to 
a bunch of accelerated H-like ions. 
The laser energy is chosen to excite the ground state $\mathrm{1s}_{1/2}$
to the excited state $\mathrm{3d}_{5/2}(m=5/2)$.
When the excited state goes back to the ground state via the electric 
quadrupole (E2) transition, a twisted photon of $(J_\gamma,M_\gamma)=(2,2)$ is emitted as 
described above.
The energy of the emitted twisted photons is up-converted by a factor of
$4\gamma_{ion}^2$ from the irradiated optical twisted laser,
where $\gamma_{ion}$ represents the Lorentz boost factor of ions and 
the case of $\gamma_{ion}\gg 1$ is considered.
The emitted photons are also highly collimated along the direction of
the ion's motion.

Although the expected flux of twisted photons in this method is sizable,
a potential problem is background. One of the major backgrounds is
the electric dipole (E1) transition from $\mathrm{3d}_{5/2}$ to
$\mathrm{2p}_{3/2}$. The E1 rate is much larger than the signal E2 rate
though the energy of the background photons is fairly different from
the signal. 
Another background is the E2 photon of $M_{\gamma}=1$ from $\mathrm{3d}_{5/2}(m=3/2)$.
This photon does not exhibit the phase singularity nor the dark spot along
the ion boost axis contrary to the one with $M_{\gamma}=2$. 
It is possible in principle to suppress this background by applying 
a transverse magnetic field to make the Zeeman splitting and selecting
the energy of the irradiating twisted laser beam so that the excitation to 
$\mathrm{3d}_{5/2}(m=3/2)$ does not occur.
This method of background suppression leads to more complication in
the design of the collision point of the gamma factory than usual.

% He-like ion
In this article, we present a new and efficient method of generating 
twisted gamma rays utilizing highly accelerated ($\gamma_{ion}\gg 1$)
helium-like (He-like) PSIs; 
an accelerated He-like PSI is excited to a chosen state by irradiating
two non-twisted optical lasers and emits a photon with OAM in the deexcitation
process. 
The energies of two laser photons seen by the ions are blue-shifted by 
a factor of $2\gamma_{ion}$, and then the energy of photons emitted in
the backward (ion-boost) direction is $2\gamma_{ion}$ times the chosen
level splitting of the ion at rest.
A salient feature of the proposed method is to utilize a decay process 
from an excited state with an angular momentum $J=2,\ M=\pm 2$
back to the ground state of $J=0,\ M=0$.
In order to reach the excited state with $J=2,\ M=\pm 2$, 
two-photon STIRAP (STImulated Raman Adiabatic Passage) \cite{Vitanov2017} 
is employed.
We examine backgrounds and the ion loss by photo-ionization as
well as the signal rate, and argue that they are not harmful.

% contents
This paper is organized as follows.
In the next section, we present a basic theory of the method, including 
its principle, calculation of ion properties and STIRAP method.  
The goal in this section is to provide the framework of the temporal evolution
of a He-like PSI under the irradiation of laser pulses.
In section 3, we present the results of our calculations, and discuss about 
signal, backgrounds, and the photoionization.
In section 4, summary is given together with discussions.
Some details are relegated to the appendix.

Throughout this paper, we use the SI unit system unless otherwise noted, and as usual,  
$c, \, \hbar,\,e$ and $m_e$ denote the light velocity in vacuum, 
the reduced Plank constant, the unit charge, and the electron mass, respectively.

%--------------Sec 2--------------------%
\section{Theory}
%--------------Sec 2--------------------%
%--------------Subsec 2.1--------------------%
\subsection{Principle of generating twisted gamma-rays}
\begin{figure*}[htb]
\begin{center}
       \includegraphics[clip,bb=30 0 750 520, width=6.5cm]{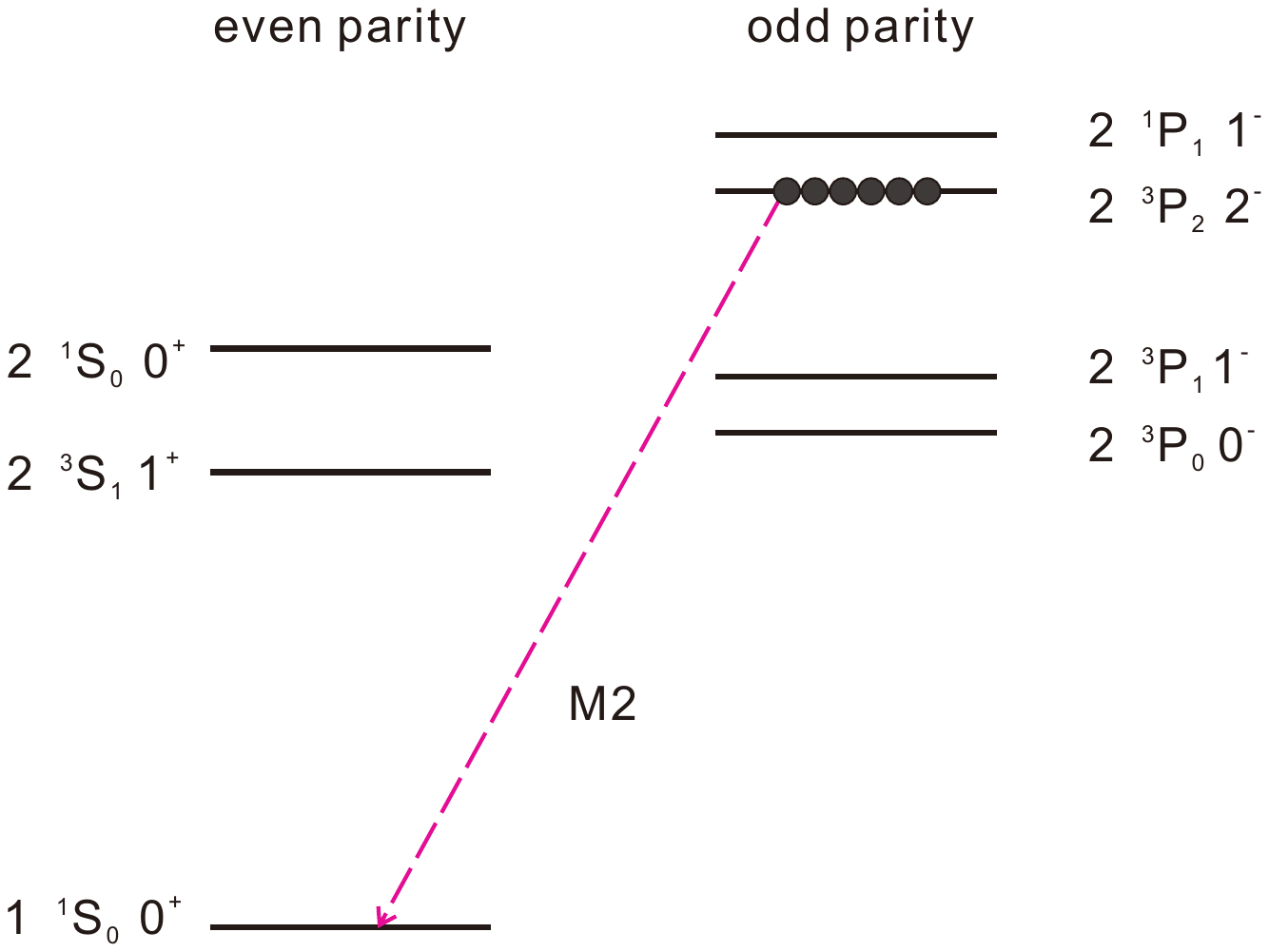}
       \caption{Schematic diagram of He-like ion energy levels. Symbols of LS-coupling scheme, 
              $n ^{2S+1}L_{J}\;J^{\pm}$, are used. 
       Populations in $2\; ^{3}P_{2}$ levels decay to the ground state via M2 transition.  Not to scale.}
       \label{fig:He-like-Energy-Diagram}
\end{center}
\end{figure*}

%--------------Subsec 2.1--------------------%
As stated already, we utilize high-energy He-like ions as an energy converter. 
For clarity, we assume that the ions are accelerated and/or stored in a circular ring 
up to the Lorentz boost factor of $\gamma_{ion}$.
A schematic energy diagram of He-like ions is shown in \Fref{fig:He-like-Energy-Diagram}. 
Let's suppose that ions are populated in a $\ket{2\,^{3}P_{2}}$ state 
with a total angular momentum $J=2$ and its magnetic quantum number $M=2$ along the beam direction. 
Here the state is specified by a standard symbol of the LS-coupling scheme, $^{2S+1\!}L_{J}$, and  
the principal quantum number $n=2$ attached in front.   
They decay back to the ground state by emitting a photon with an angular momentum of 2. 
For a sufficiently large $\gamma_{ion}$, say $>1000$, the energy of emitted photon is amplified to a 
gamma ray region in a laboratory.   
One crucial feature of He-like ions is that the transition from $\ket{2\; ^{3}P_{2}}$ to the ground state, 
magnetic quadrupole (M2) in nature, is a major decay channel.  
This is due to a relativistic effect: the M2 transition is expected to scale with $Z^8$, which 
becomes sizable when high $Z$ ions are selected \cite{Sobelman,Lin1977}. 
The STIRAP method, 
in which two circularly polarized lasers are irradiated to populate the $J=2$ and $M=2$ states,  
is used for an efficient transition. 
This is the basic principle of generating gamma-rays with OAM.
Below we consider He-like Kr ($Z=36$) with $\gamma_{ion}=2500$ and Xe ($Z=54$) with $\gamma_{ion}=5000$ as an illustration. 
Gamma-rays with energies of $\sim 65$ MeV (Kr) or $\sim 306$ MeV (Xe) are obtained for $\gamma_{ion}$ chosen above.
Similar discussions and/or conclusions apply equally to other He-like ions.

%
%+++++++++++++++++++++++++++++ figures ++++++++++++++++++++++++++++++
%+++++++++++++++++++++++++++++ figures ++++++++++++++++++++++++++++++
%

%--------------Subsec 2.2--------------------%
\subsection{Properties of He-like ions--Details of calculations--}
%--------------Subsec 2.2--------------------%
%\subsubsection{Details of calculations}
\subsubsection*{Energy levels and radiative transitions\\}
\begin{figure*}[htb]
\begin{center}
\begin{minipage}{7.5cm}
	\includegraphics[angle=0,clip,bb=50 0 750 550, width=7.5cm]{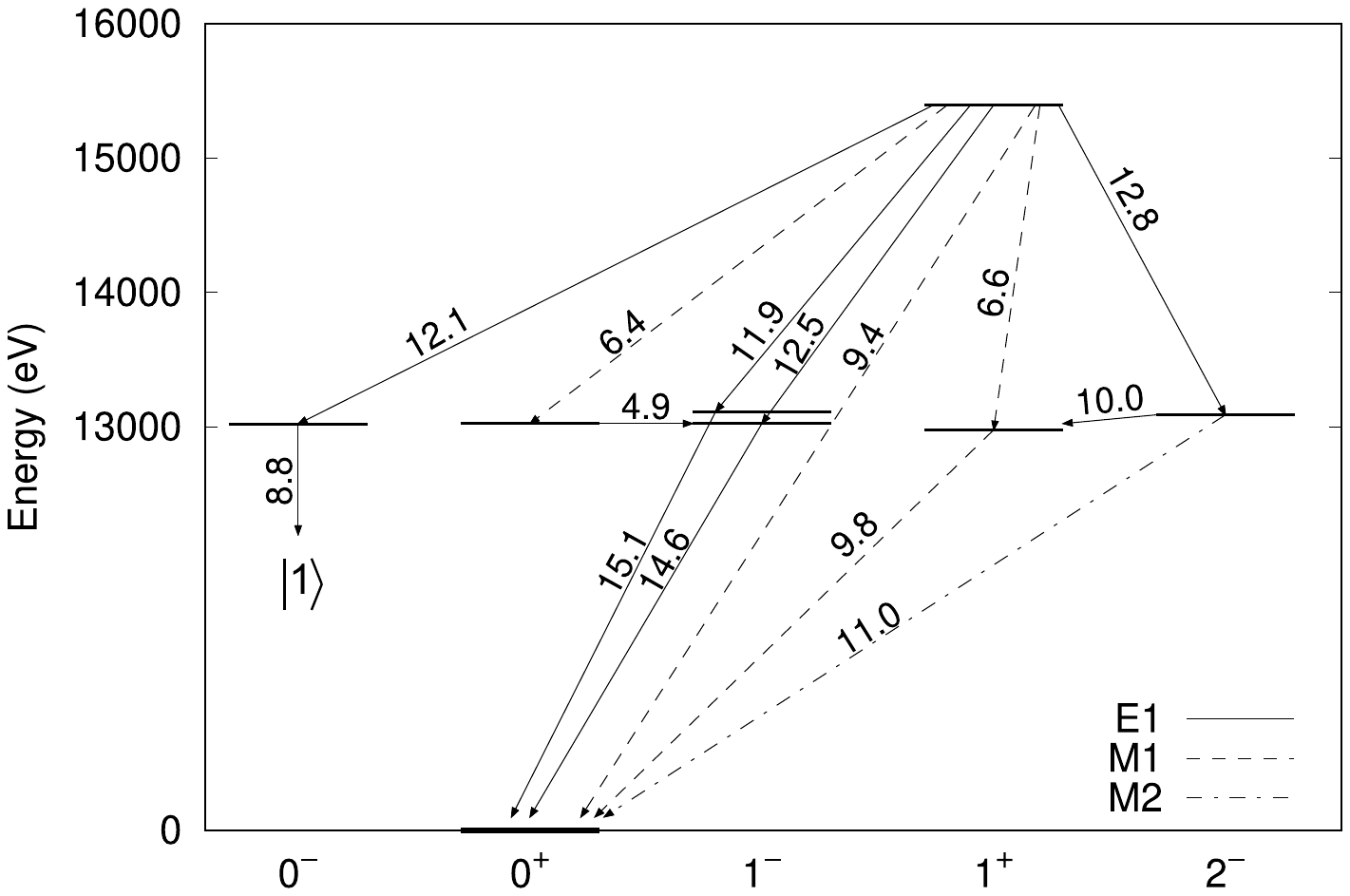}
\end{minipage}
\begin{minipage}{7.5cm}
	\includegraphics[angle=0,clip,bb=50 0 750 550, width=7.5cm]{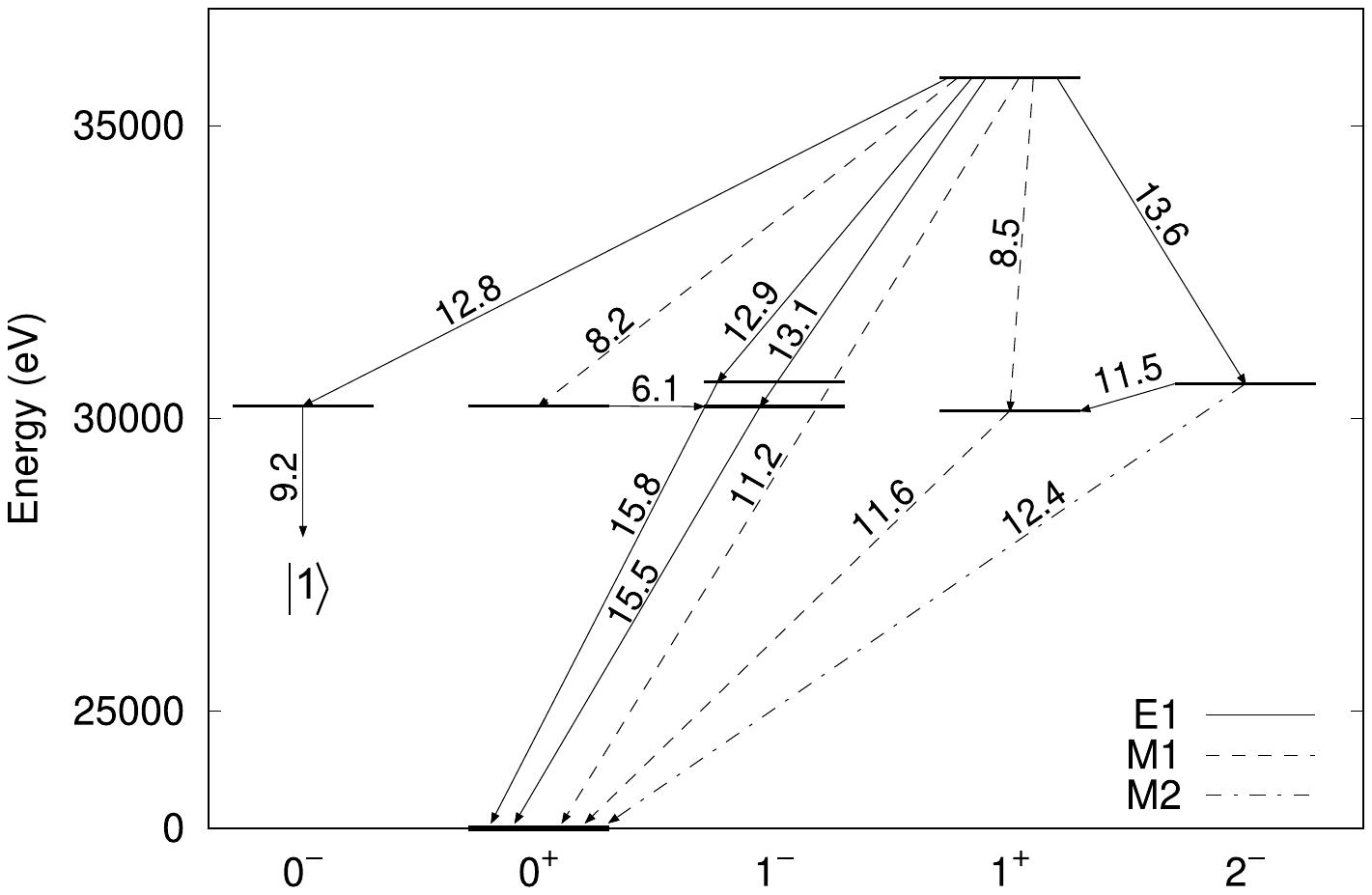}
\end{minipage}
\end{center}
\caption{Low-lying energy levels of He-like Kr (left) and Xe (right). 
       Major radiative transitions are indicated by arrows 
       (E1 by solid,  M1 by dashed, and M2 by dash-dotted lines) 
       together with A-coefficients (powers of 10).}
       \label{fig:Theory-Xe-Kr}
\end{figure*}

In this work, energy levels of He-like Kr and Xe were calculated by
the multi-configuration Dirac-Hartree-Fock method and the relativistic configuration interaction
method \cite{grant} implemented in the GRASP2018 package \cite{grasp2018}.
The wave functions for these atomic states were represented by linear combinations of 
configuration state functions, which were constructed from single-particle Dirac orbitals. 
The single-particle Dirac orbitals were determined by the multi-configuration 
Dirac-Hartree-Fock method, whereas the expansion coefficients of the linear combination 
were calculated by the relativistic configuration interaction method. 
Using the multi-configuration Dirac-Hartree-Fock method,  single-particle Dirac orbitals were obtained up to $7s$,$7p$,$7d$,$6f$ and $6g$. Treating these orbitals as active space, 
the relativistic configuration interaction calculations were performed to include the Breit interaction \cite{grant}, vacuum polarization \cite{vac pol},
and self-energy \cite{vac pol} effects. Total of 17 electronic states as well as  
the amplitudes of the E1, E2, M1 and M2 transitions among these states were calculated 
in this work.
The results of calculations are summarized in 
\Tref{table:all states}, and are illustrated in \Fref{fig:Theory-Xe-Kr}.
%
%--------------------------------Table-------------------------------------
\begin{longtable}{rl|rc|cc} 
\hline
  & &  \multicolumn{2}{c|}{He-like Kr}    & \multicolumn{2}{c}{He-like Xe} \\
    Label & State                     & Energy /eV  & $\tau$ /s & Energy /eV  & $\tau$ /s\\ \hline
$\ket{g}$ & $1~0^+~$$1s(0)~^1S_{  0  }$      & 0.000      &  -                   & 0.000 &  -  \\
$\ket{1}$ & $1~1^+~$$1s\,2s~^3S_{  1  }$     & 12978.4 &  1.749$\times 10^{-10}$  & 30127.3 &  2.627$\times 10^{-12}$\\
$\ket{5}$ & $1~0^-~$$1s\,2p~^3P_{  0  }^o$   & 13022.5 &  1.456$\times 10^{-9}$   & 30210.3 &  5.836$\times 10^{-10}$\\
$\ket{3}$ & $1~1^-~$$1s\,2p~^3P_{  1  }^o$   & 13025.2 &  2.540$\times 10^{-15}$  & 30204.3 &  3.270$\times 10^{-16}$\\
$\ket{4}$ & $2~0^+~$$1s\,2s~^1S_{  0  }$     & 13025.9 &  1.173$\times 10^{-5}$   & 30212.8 &  8.098$\times 10^{-7}$\\
$\ket{f}$ & $1~2^-~$$1s\,2p~^3P_{  2  }^o$   & 13089.9 &  9.571$\times 10^{-12}$  & 30592.8 &  3.458$\times 10^{-13}$\\
$\ket{2}$ & $2~1^-~$$1s\,2p~^1P_{  1  }^o$   & 13113.7 &  6.551$\times 10^{-16}$  & 30628.0 &  1.471$\times 10^{-16}$\\
$\ket{e}$ & $2~1^+~$$1s\,3s~^3S_{  1  }$     & 15394.4 &  8.400$\times 10^{-14}$  & 35821.1 &  1.396$\times 10^{-14}$\\
          & $3~0^+~$$1s\,3s~^1S_{  0  }$     & 15407.0 &  8.929$\times 10^{-14}$  & 35843.8 &  1.462$\times 10^{-14}$\\
          & $2~0^-~$$1s\,3p~^3P_{  0  }^o$   & 15407.1 &  2.846$\times 10^{-14}$  & 35844.3 &  1.008$\times 10^{-15}$\\
          & $3~1^-~$$1s\,3p~^3P_{  1  }^o$   & 15407.8 &  6.969$\times 10^{-15}$  & 35845.8 &  5.077$\times 10^{-15}$\\
          & $2~2^-~$$1s\,3p~^3P_{  2  }^o$   & 15427.1 &  3.038$\times 10^{-14}$  & 35959.6 &  5.937$\times 10^{-15}$\\
          & $1~2^+~$$1s\,3d~^3D_{  2  }$     & 15433.0 &  9.942$\times 10^{-15}$  & 35968.3 &  1.818$\times 10^{-15}$\\
          & $3~1^+~$$1s\,3d~^3D_{  1  }$     & 15433.4 &  9.986$\times 10^{-15}$  & 35969.7 &  4.821$\times 10^{-16}$\\
          & $4~1^-~$$1s\,3p~^1P_{  1  }^o$   & 15434.6 &  2.232$\times 10^{-15}$  & 35969.8 &  1.850$\times 10^{-15}$\\
          & $1~3^+~$$1s\,3d~^3D_{  3  }$     & 15439.7 &  1.023$\times 10^{-14}$  & 36004.5 &  1.955$\times 10^{-15}$\\
          & $2~2^+~$$1s\,3d~^1D_{  2  }$     & 15440.3 &  1.019$\times 10^{-14}$  & 36005.9 &  1.915$\times 10^{-15}$\\
\hline\\
\caption{Low-lying levels of He-like Kr and Xe, and their energies and lifetimes. 
Ascending order in energy except $\ket{3}$ and $\ket{5}$ of Xe. 
The states $\ket{i}\,(i=g,e,f,1 \cdots 5)$ are considered in the simulation study. 
}
\label{table:all states}
\end{longtable}
%--------------------------------Table-------------------------------------

%
%+++++++++++++++++++++++++++++ figures ++++++++++++++++++++++++++++++
%+++++++++++++++++++++++++++++ figures ++++++++++++++++++++++++++++++
%

\subsubsection*{Photo-ionization cross section\\[2mm]}
To estimate the fraction of ionization caused by the laser used in the experiment, 
photoionization cross sections of He-like ions were calculated by using the RATIP 
programs\cite{ratip}. The wavefunctions of the ions used in the photoionization 
calculation were basically the same as described just above. 
The results of calculations,  relevant to our present studies, are summarized in \Tref{table:Photo-ionization Cross-section}.

%---------------------Table-------------------------%
\begin{table}[h]
\caption{Photo-ionization Cross-section}
\begin{center}
\begin{tabular}{cc|rr|rr} \hline
  & &  \multicolumn{2}{c|}{He-like Kr}    & \multicolumn{2}{c}{He-like Xe} \\
Symbol & Initial state   &  $\hbar \omega$/eV & $\sigma^{(pi)}$/barn &  $\hbar \omega$/eV& $\sigma^{(pi)}$/barn\\ \hline
$\sigma_{es}^{(pi)}$ & $2~1^+~$$1s\,3s~^3S_{  1  }$  & 2304.5 & 15330 & 5228.3 & 6839 \\
$\sigma_{ep}^{(pi)}$ & $2~1^+~$$1s\,3s~^3S_{  1  }$  & 15394.4 & 203.0  & 35821.1 & 84.78 \\
$\sigma_{fp}^{(pi)}$ & $1~2^-~$$1s\,2p~^3P_{  2  }^o$ & 15394.4 & 90.78  & 35821.1 & 36.60 \\
$\sigma_{1p}^{(pi)}$ & $1~1^+~$$1s\,2s~^3S_{  1  }$ & 15394.4 & 709.7  & 35821.1 & 292.8 \\ 
$\sigma_{2p}^{(pi)}$ & $2~1^-~$$1s\,2p~^1P_{  1  }^o$ & 15394.4 & 139.2  & 35821.1 & 57.22 \\ 
$\sigma_{3p}^{(pi)}$ & $1~1^-~$$1s\,2p~^3P_{  1  }^o$ & 15394.4 & 148.4  & 35821.1 & 70.74 \\
$\sigma_{4p}^{(pi)}$ & $2~0^+~$$1s\,2s~^1S_{  0  }$ & 15394.4 & 747.1 & 35821.1 & 311.2 \\
$\sigma_{5p}^{(pi)}$ & $1~0^-~$$1s\,2p~^3P_{  0  }^o$ & 15394.4 & 206.6  & 35821.1 & 97.29 \\
\hline
\end{tabular}
\label{table:Photo-ionization Cross-section}
\end{center}
\end{table}
%---------------------Table-------------------------%

%--------------Subsec 2.3--------------------%
\subsection{Population transfer by Raman process}\label{sec:Population transfer by Raman}
\begin{figure*}[htb]
\begin{center}
       \includegraphics[clip,bb=0 0 900 600, width=8.0cm]{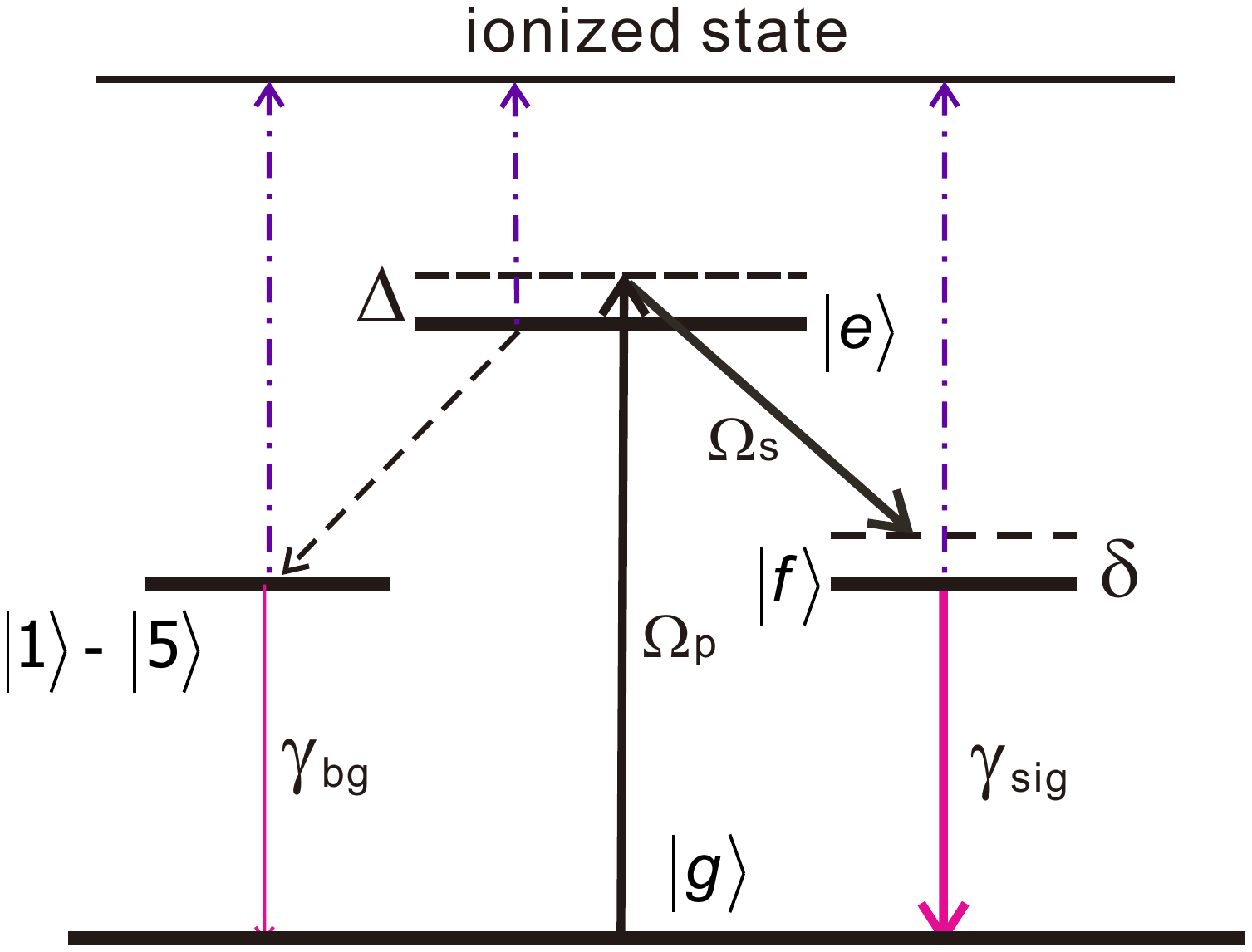}
       \caption{Levels and processes relevant to the STIRAP simulation. The major levels are 
       $\ket{g}=\ket{1\, ^{1\!}S_{0}}$, $\ket{e}=\ket{3\, ^{3\!}S_{1}}$, and $\ket{f}=\ket{2\, ^{3\!}P_{2}}$. 
       The thick solid arrows indicate the main path for generating $\gamma_{sig}$.
       $\Omega_p (\Omega_s)$ denotes Rabi frequency of the pump (Stokes) laser, and $\Delta$ ($\delta$) is 
       pump laser (two-photon) detuning. 
       The dashed line represents the spontaneous radiative decays and 
       the dash-dotted ones indicate photo-ionization processes. 
       See \Tref{table:all states} for the levels labeled by Arabic numerals.
       }
       \label{fig:STIRAP-Diagram}
\end{center}
\end{figure*}

%--------------Subsec 2.3--------------------%
The technique of stimulated Raman adiabatic passage (STIRAP) is now a well established one: 
it allows efficient and selective population transfer between quantum states without suffering loss 
due to spontaneous emission \cite{Vitanov2017}. 
As illustrated by the thick solid arrows in \Fref{fig:STIRAP-Diagram}, 
two lasers (pump and Stokes) are irradiated to stimulate transitions from 
the ground state $\ket{g}=\ket{1\, ^{1\!}S_{0}}$ to 
the final state $\ket{f}=\ket{2\, ^{3\!}P_{2}}$ via 
the excited state $\ket{e}=\ket{3\,^{3\!}S_{1}}$. 
The Hamiltonian for the process is given by
\begin{eqnarray}
 && H(t)=\hbar 
        \left[ \begin{array}{ccc} 
 	0& \frac{1}{2}\Omega_P & 0\\
        \frac{1}{2}\Omega_P^{\ast} & -\Delta &  \frac{1}{2}\Omega_S^{\ast}\\ 
        0 & \frac{1}{2}\Omega_S &  -\delta  \\
        \end{array} \right]
 \label{eq:Basic formula} 
\end{eqnarray}
after applying the rotating wave approximation \cite{Vitanov2017}.  
Here $\displaystyle \Omega_P$ and $\displaystyle \Omega_S$ 
denote Rabi frequencies of pump and Stokes laser fields, respectively, 
and $\Delta\; (\delta)$ is the pump laser detuning (two-photon detuning) defined as 
\begin{eqnarray}
 && \Delta \equiv\omega_p- \omega_{eg}, \hspace{10mm}
 \delta\equiv \omega_p-\omega_s-\omega_{fg} 
 \label{eq:def detuning} 
\end{eqnarray}
where $\omega_p$ ($\omega_s$) is the pump (Stokes) laser frequency, and 
$\hbar \omega_{eg}$ ($\hbar \omega_{fg}$) is the $\ket{e}-\ket{g}$ ($\ket{f}-\ket{g}$) energy difference.
The Rabi frequencies are calculated with respective A-coefficients and 
are specified in more detail below. 
In addition to the Raman process, processes such as spontaneous decay or photo-ionization are of practical importance. 
To study their effects, all $n=2$ states, labeled as $\ket{1}\sim \ket{5}$ in \Fref{fig:STIRAP-Diagram},
are considered in the simulation described below. See \Tref{table:all states} for the actual assignment.
These states are populated by the spontaneous emission from the excited state $\ket{e}$ 
(and less importantly from higher levels within $n=2$), 
and generate background gamma-rays with energies similar to the signal 
by decaying to the ground state $\ket{g}$.

%+++++++++++++++++++++++++++++ figures ++++++++++++++++++++++++++++++
%+++++++++++++++++++++++++++++ figures ++++++++++++++++++++++++++++++
%

\paragraph{Optical Bloch equations\\}
In order to examine population transfer efficiency, we employ optical Bloch equations, 
in which effects of all major decays are taken into account based on the master equation\cite{lindblad1976}.
The optical Bloch equations (in the ion-rest frame) are given by  
\begin{eqnarray}
	&&\frac{d\rho_{gg}}{dt}= -\Re\Big( i \Omega_{p} \rho_{ge}^{\ast} \Big) 
             + \sum_{j=e,f}^{1\sim5}\Gamma_{jg}\rho_{jj}
                \nonumber \\
        &&\frac{d\rho_{ee}}{dt}=+\Re\Big( i \Omega_{p} \rho_{ge}^{\ast} + i\Omega_{s} \rho_{fe}^{\ast} \Big)
        	 -\Big(\Gamma_{e}^{(tot)}+\Gamma_{e}^{(pi)}\Big)\rho_{11}
                 \nonumber \\
        &&\frac{d\rho_{ff}}{dt}= -\Re\Big( i\Omega_{s} \rho_{fe}^{\ast} \Big)+ \Gamma_{ef}\rho_{ee}
        -\Big(\Gamma_{f}^{(tot)}+\Gamma_{f}^{(pi)}\Big)\rho_{ff}
        \nonumber \\
        &&\frac{d\rho_{ge}}{dt}+i \Delta \rho_{ge}=\frac{i}{2} \Big( \Omega_{s}\rho_{gf} 
        	+ \Omega_{p}(\rho_{gg}-\rho_{ee}) \Big)-\gamma_{ge}\rho_{ge}
                \nonumber \\
        &&\frac{d\rho_{gf}}{dt}+i \delta \rho_{gf}=\frac{i}{2} \Big( 
		\Omega_{s}^{\ast}\rho_{ge} - \Omega_{p}\rho_{fe}^{\ast} \Big)-\gamma_{gf}\rho_{gf}
        \nonumber \\        
                &&\frac{d\rho_{fe}}{dt}+i (\Delta-\delta) \rho_{fe} =\frac{i}{2} 
                \Big(\Omega_{p}\rho_{gf}^{\ast} + \Omega_{s}(\rho_{ff}-\rho_{ee}) \Big)-\gamma_{fe}\rho_{fe}
        \nonumber \\
        &&\frac{d\rho_{ii}}{dt}=\Big(\Gamma_{ei}\rho_{ee}+\Gamma_{ji}\rho_{jj}\Big)
        -\Big(\Gamma_{ig}+\Gamma_{ij}+\Gamma_{i}^{(pi)}\Big)\rho_{ii}, 
        \hspace{4mm}(i,j)=(1,\cdots,5)
        \label{eq:1-Level-Optical-Bloch-with-relaxation-terms} 
\end{eqnarray}
where $\rho_{ii}\;(\rho_{ij})$ denotes the population of the state $\ket{i}$ 
(coherence between $\ket{i}$ and $\ket{j}$), and 
$\Gamma_{i}^{(tot)}\;(\Gamma_{ij})$ is the corresponding total (partial) width. 
The actual values of $1/\Gamma_{i}^{(tot)}$({\it i.e.} lifetime) and $\Gamma_{ij}$ are specified 
in \Tref{table:all states} or in \Tref{table:Radiative transitions}.  
Using \Eref{eq:E1 and M1 Rabi} in Appendix, 
the Rabi frequencies are given by 
\begin{eqnarray}
 && \Omega_{p}(t)=\frac{e E_{p}(t)}{\hbar}\sqrt{\frac{3\,\Gamma_{eg}}{4\, c \alpha k_p^3}}, 
 %\hspace{5mm}\mathrm{and}\hspace{5mm}
 \nonumber \\ &&
 \Omega_{s}(t)=\frac{e E_{s}(t)}{\hbar}\sqrt{\frac{9\,\Gamma_{ef}}{20\, c \alpha k_s^3}},
 \label{eq:Rabi Omega} 
\end{eqnarray}
where $E_{p}$ ($E_{s}$) denotes the magnitude of the pump (Stokes) laser field with 
wavenumber $k_p$ ($k_s$) and $\alpha\simeq1/137$ is the fine structure constant.  
Table \ref{table:Rabi frequency} shows the Rabi frequencies for an unit input power $I_{in}=1$ W/mm$^2$. 
As seen, there is a large imbalance between $\Omega_p$ and $\Omega_s$. 

The transverse decay (decoherence) are assumed to stem from longitudinal components 
and their constants are taken as follows:
\begin{eqnarray}
 && \gamma_{ge}=\frac{1}{2}\Gamma_{e}^{(tot)}, \hspace{10mm}
 \gamma_{gf}=\frac{1}{2}\Gamma_{f}^{(tot)}
 \nonumber \\ 
 && \gamma_{fe}=\frac{1}{2}\left( \Gamma_{e}^{(tot)}+\Gamma_{f}^{(tot)}\right).
 \label{eq:decoherence terms} 
\end{eqnarray}
The terms $\Gamma_{i}^{(pi)}\;(i=e,f,1,\cdots,5)$ need special attention. 
These terms represent a photo-ionization effect due to absorption of additional laser photons. 
The ionization rate is expressed by 
\begin{eqnarray}
 &&  \Gamma_{e}^{(pi)}=\sigma^{(pi)}_{es}\frac{I_{s}(t)}{\hbar \omega_{s}}+
 \sigma^{(pi)}_{ep}\frac{I_{p}(t)}{\hbar \omega_{p}},  
 \nonumber \\
 && \Gamma_{i}^{(pi)}=\sigma^{(pi)}_{ip}\frac{I_{p}(t)}{\hbar \omega_{p}}\hspace{10mm} (i=f,1,\cdots,5),
 \label{eq:photo-ionization rate Gamma26 Gamma36} 
\end{eqnarray}
where $\sigma^{(pi)}_{ip}$ or $\sigma^{(pi)}_{is}$ denotes the photo-ionization cross-section of the state $\ket{i}$ 
by pump or Stokes lasers, 
$I_{p,s}(t)$ the laser intensity, and  
$\displaystyle \frac{I_{p,s}(t)}{\hbar \omega_{p,s}}$ the photon number flux ($\omega_{p,s}=ck_{p,s}$).
We note that the energy of Stokes laser photons is below the photo-ionization threshold except for $\ket{e}$. 
We also note that since $\ket{e}$ is the lowest among the $n=3$ states, 
there is no decay path going through them.  

%---------------------Table-------------------------%
\begin{table}[h]
\caption{Rabi frequency for unit input power $I_{in}=1$ W/mm$^2$ (in the ion-at-rest frame). }
\begin{center}
\begin{tabular}{cccccc} \hline
Transitions & Type  &   Kr [s$^{-1}$]   & Xe [s$^{-1}$]  \\ \hline
$\Omega_p\;\;$ $\ket{g}\to \ket{e}$ & M1 & $5.570 \times 10^4$ & $1.276 \times 10^5$  \\
$\Omega_s\;\;$ $\ket{e}\to \ket{f}$ & E1 & $3.935 \times 10^7$ & $2.882 \times 10^7$  \\
\hline
\end{tabular}
\label{table:Rabi frequency}
\end{center}
\end{table}
%---------------------Table-------------------------%

%---------------------Table-------------------------%
\begin{table}[h]
\caption{Radiative transitions  used in the simulation}
\begin{center}
\begin{tabular}{cc|cc|cc} \hline
  &     & \multicolumn{2}{c|}{He-like Kr}  &  \multicolumn{2}{c}{He-like Xe} \\
Transitions & Type  &   Energy/eV  & $\Gamma$/s$^{-1}$ & Energy/eV  & $\Gamma$/s$^{-1}$ \\ \hline
$\ket{e}\to \ket{g}$ & M1 & 15394 & $2.469 \times10^{9}$ & 35821 & $1.633\times10^{11}$ \\
$\ket{e}\to \ket{f}$ & E1 & 2305 & $6.891 \times 10^{12}$ & 5228  & $4.317 \times 10^{13}$\\
$\ket{e}\to \ket{1}$ & M1 & 2416 & $4.066 \times 10^{6}$ & 5694 & $2.936 \times 10^{8}$\\
$\ket{e}\to \ket{2}$ & E1 & 2281 & $8.444 \times 10^{11}$ & 5193 & $7.705 \times 10^{12}$ \\
$\ket{e}\to \ket{3}$ & E1 & 2369 & $2.927 \times 10^{12}$ & 5617 & $1.388 \times 10^{13}$ \\
$\ket{e}\to \ket{4}$ & M1 & 2369 & $2.289 \times 10^{6}$ & 5608 & $1.603 \times 10^{8}$ \\
$\ket{e}\to \ket{5}$ & E1 &  2372 & $1.241 \times 10^{12}$ & 5611 & $6.701 \times 10^{12}$ \\
$\ket{f}\to \ket{g}$ & M2 &  13090 & $9.309 \times 10^{10}$ & 30593 & $2.554 \times 10^{12}$ \\
$\ket{1}\to \ket{g}$ & M1 &  12978 & $5.719 \times 10^{9}$ & 30127 & $3.806 \times 10^{11}$ \\
$\ket{2}\to \ket{g}$ & E1 &  13114 & $1.526 \times 10^{15}$ & 30628 & $6.798 \times 10^{15}$ \\
$\ket{3}\to \ket{g}$ & E1 &  13025 & $3.936 \times 10^{14}$ & 30204 & $3.058 \times 10^{15}$ \\
$\ket{f}\to \ket{1}$ & E1 &  112 & $1.139 \times 10^{10}$ & 465 & $3.374 \times 10^{11}$ \\
$\ket{4}\to \ket{3}$ & E1 &  13026 & $8.385 \times 10^{4}$ & 30213 & $1.192 \times 10^{6}$ \\
$\ket{5}\to \ket{1}$ & E1 &  0.7 & $6.866 \times 10^{8}$ & 8.5 & $1.714 \times 10^{9}$ \\
\hline
\end{tabular}
\label{table:Radiative transitions}
\end{center}
\end{table}
%---------------------Table-------------------------%

%--------------Sec 3--------------------%
\section{Simulation Results}
%--------------Sec 3--------------------%
%
%--------------Subsec 3.1--------------------%
\subsection{Overview of the simulation results\\}\label{sec:sec31}
\begin{figure}[htb]
\begin{center}
\begin{minipage}{7.5cm}
       \includegraphics[clip,bb=0 0 380 260, width=7.5cm]{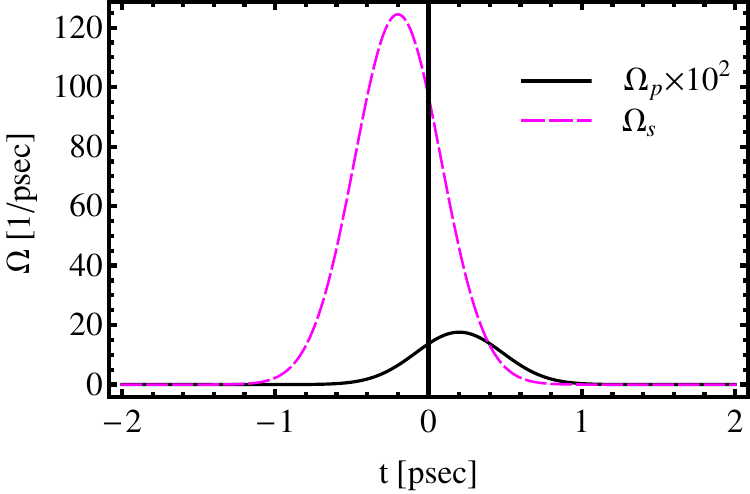}
\end{minipage}
\begin{minipage}{7.5cm}
       \includegraphics[clip,bb=0 0 380 260, width=7.5cm]{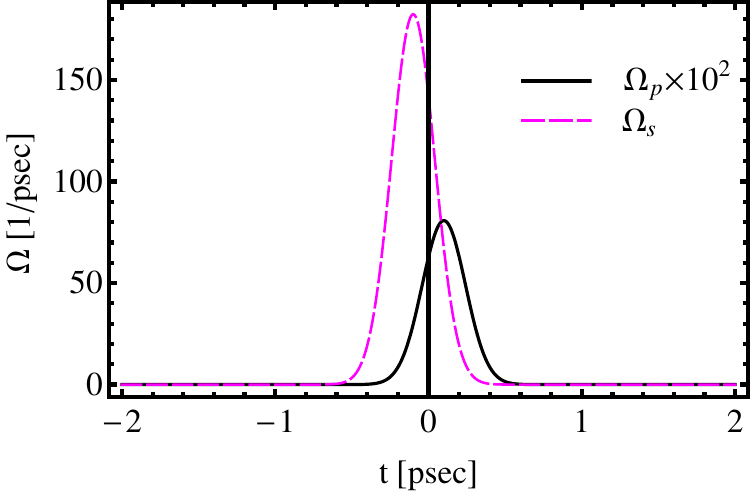}
\end{minipage}
       %\captionsetup{width=.9\linewidth}
       \caption{Rabi frequency profiles vs time $t$ in the ion-at-rest frame  
       for Kr (left) and Xe (right).
       The pump ($\times 10^{2}$, in black solid) and Stokes (in red dashed).
       See Table \ref{list of parameters sets} for the parameters used.}
       \label{fig:fig00Rabi}
\end{center}
\end{figure}

%--------------Subsec 3.1--------------------%
We first describe a basic parameter set used in the simulation.  
See Table \ref{list of parameters sets} (note that values are all in the laboratory frame).
The Lorentz boost factor $\gamma_{ion}$ is assumed to be $\gamma_{ion}=2500$ for Kr and 
$\gamma_{ion}=5000$ for Xe. 
These choices are made so that wavelengths of input lasers fall into a convenient region (visible or infrared).
Assuming that an ion beam has a bunching time structure in a ring accelerator, 
we use pulsed lasers for both pump and Stokes.   
The time profiles are assumed to be Gaussian, and are expressed by 
\begin{eqnarray}
 && I_{p,s}(t)=I_{p,s}(0) \exp\left(-\frac{(t\mp t_d)^2}{2 \sigma_L^2}\right),
 \label{eq:Laser pulse shape} 
\end{eqnarray}
where $\sigma_L$ denotes a root-mean-square width and $t_{d}$ a time difference between the two pulses 
 (the Stokes precedes the pump by $2t_{d}$).
We call $t_{d}$ a laser delay time for convenience. 
The width $\sigma_L$ and delay time $t_{d}$ are taken to be 1 nsec and 0.5 nsec, respectively. 
The peak intensity $I_p(0)$ and $I_s(0)$ are assumed to be equal and are 400 kW/mm$^2$. 
With the laser spot size of 1 mm$^2$, 
the laser's total pulse energies amount to 1 mJ per pulse for each. 
The pump laser detuning $\Delta$ is taken to be 5 times $\Gamma_{e}^{(tot)}$, and  
the two-photon detuning is set to $\delta=0$.
The Rabi frequency profiles, calculated according to \Eref{eq:Rabi Omega} 
 are shown in \Fref{fig:fig00Rabi}. 
In the following section, we assume the transverse ion beam size matches with the laser spot size 
({\it i.e.} less than 1 mm$^2$).

%---------------------Table-------------------------%
\begin{table}[h]
\caption{Basic parameter set used in the simulation. }
\begin{center}
\begin{tabular}{rcccl} \hline
Parameter & Symbol    & Kr  & Xe  & units  \\
\hline
Lorentz boost factor & $\gamma_{ion}$ & 2500 & 5000 & - \\
Pump laser wavelength & $\lambda_p$  & 403 & 346 & nm\\
Pump laser intensity & $I_p(0)$  & $4\times 10^5$ &$4\times 10^5$ & W/mm$^2$\\
Stokes laser wavelength & $\lambda_s$ &  2690 & 2371 & nm  \\
Stokes laser intensity& $I_s(0)$ & $4\times 10^5$ & $4\times 10^5$  &W/mm$^2$\\
Laser pulse width & $\sigma_L$  & 1.0 & 1.0 & nsec \\ 
Laser pulse delay & $t_d$ &  $0.5$ & $0.5$ & nsec \\
Pump laser detuning  & $\Delta$ & 5 & 5 & $\Gamma_{e}^{(tot)}$   \\
Two-photon detuning  & $\delta$ & 0 & 0 & $\Gamma_{e}^{(tot)}$   \\
\hline 
\end{tabular}
\\[2mm]
(Values are all in the laboratory frame.)
\label{list of parameters sets}
\end{center}
\end{table}
%---------------------Table-------------------------%

%+++++++++++++++++++++++++++++ figures ++++++++++++++++++++++++++++++
%+++++++++++++++++++++++++++++ figures ++++++++++++++++++++++++++++++

%--------------Subsec 3.2--------------------%
\subsection{Main simulation results\\}
\begin{figure}[htb]
\begin{center}
\begin{minipage}{7.5cm}
       \includegraphics[clip,bb=0 0 380 260, width=7.5cm]{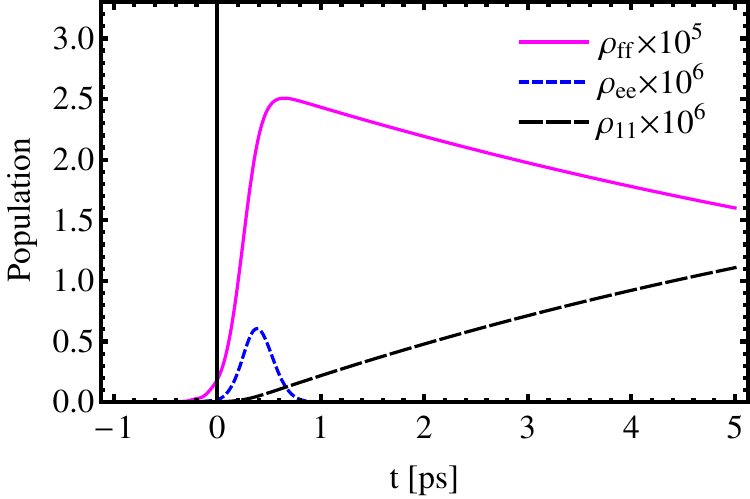}
\end{minipage}
\begin{minipage}{7.5cm}
\includegraphics[clip,bb=0 0 380 260, width=7.5cm]{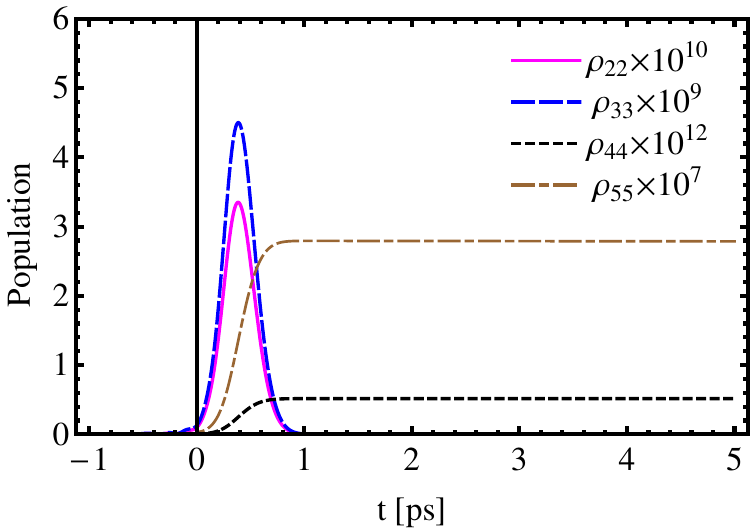}
\end{minipage}
       %\captionsetup{width=.9\linewidth}
       \caption{Variation of populations as a function of time $t$ in the ion-at-rest frame (Kr). 
       (left) $\rho_{ff}\times10^{5}$ (solid magenta),    
       $\rho_{11}\times 10^{6}$ (dashed black),  
       $\rho_{ee}\times 10^{6}$  (dotted blue).
       (right) $\rho_{22}\times10^{10}$ (solid magenta),
       $\rho_{33}\times 10^{9}$ (dashed blue),
       $\rho_{44}\times 10^{12}$ (dotted black), 
       $\rho_{55}\times 10^{7}$ (dash-dotted brown).
       See Table \ref{list of parameters sets} (Kr) for the parameters used. 
       }
       \label{fig:Kr-population}
\end{center}
\end{figure}

%--------------Subsec 3.2--------------------%
We now show our simulation results using Kr ions as a prime example. 
\Fref{fig:Kr-population} (left) shows time variation of populations of the excited state $\ket{e}$, 
the final state $\ket{f}$, and the state $\ket{1}=\ket{2^{3\!}S_1}$. 
Note that $t$ in this figure (and also the following figures) represents the time in the ion-at-rest frame. 
As seen, a sizable fraction is transported to $\ket{f}$ although excitation to  $\ket{e}$ is non-negligible. 
The intermediate state $\ket{1}$ increases gradually because it is mainly generated through $\ket{f}$. 
See \Fref{fig:Theory-Xe-Kr} and/or \Tref{table:Radiative transitions}.
The decay from $\ket{f}$ to the ground state $\ket{g}$ occurs quickly, and gives the twisted gamma-rays (signal events). 
\Fref{fig:Kr-population} (right)  shows the time variation of the other populations, $\ket{2}\sim \ket{5}$: 
they are found to be at least two order of magnitude smaller than that of $\ket{f}$.

%+++++++++++++++++++++++++++++ figures ++++++++++++++++++++++++++++++
%+++++++++++++++++++++++++++++ figures ++++++++++++++++++++++++++++++

\begin{figure}[htb]
\begin{center}
       \includegraphics[clip,bb=0 0 380 260, width=7.5cm]{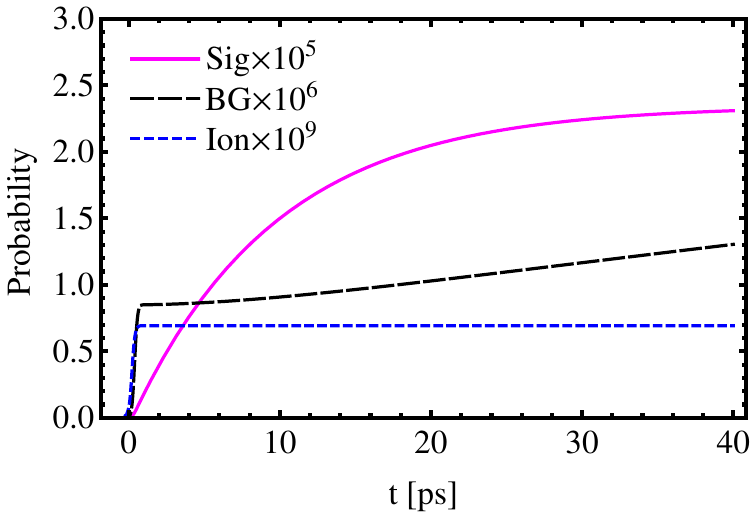}
       %\captionsetup{width=.9\linewidth}
       \caption{Event probability integrated up to time t. 
       Signal ($\times 10^5$, solid magenta), background ($\times 10^6$, dashed black), 
       and ionization ($\times 10^{9}$, dotted blue). 
       See Table \ref{list of parameters sets} (Kr) for the parameters used. 
       }
       \label{fig:Set-1-fig07}
\end{center}
\end{figure}

Now we discuss about more important quantities such as signal or background 
production probability (or ``event rate" for simplicity). 
The solid magenta line in \Fref{fig:Set-1-fig07} shows the probability 
of signal events integrated up to time $t$. 
It increases gradually with a time constant nearly equal to the lifetime of $\ket{f}$. 
Photons from the intermediate states $\ket{1}\sim \ket{5}$ decaying to the ground state $\ket{g}$ have 
similar energies with the signal, and may 
constitute undesirable background events (though it depends on actual experimental conditions).  
The dashed black line in \Fref{fig:Set-1-fig07} shows the total probability 
summed over all contributions from $\ket{1}\sim \ket{5}$ 
\footnote{A direct decay from $\ket{e}$ to $\ket{g}$ is another source of backgrounds. 
It is neglected, however, because its contribution is less than $10^{-3}$ of those mentioned above. 
In addition, the resultant $\gamma$-ray energy differs significantly from that of the signal.}. 
A step-function-like increase near $t=0$ is due the short-lived states of 
$\ket{2}=\ket{2 ^{1\!}P_1}$ and $\ket{3}=\ket{2 ^{3\!}P_1}$ while 
more gradual increase is mainly by $\ket{1}$. 
Note that since $\ket{1}$ has a longer life time than $\ket{f}$, it 
is still increasing at $t=40$ psec. If we can bend the ion beam in the accelerator quickly enough, some 
of these background events would be directed away from an experimental area. 
In this paper, we evaluate the background events (and also the signal) at $t=40$ psec:  
this choice assumes that a bending section starts at 30 m downstream of the interaction (straight) section in the laboratory.  
With this definition, the signal and background rates are about $2.3\times 10^{-5}$ and $1.3 \times 10^{-6}$, respectively, 
giving the background-to-signal ratio ($B/S$) of 5.6\%. 
%
%+++++++++++++++++++++++++++++ figures ++++++++++++++++++++++++++++++
%+++++++++++++++++++++++++++++ figures ++++++++++++++++++++++++++++++
%

We next consider a photo-ionization effect, which is potentially serious 
because once ions are ionized (forming H-like ions) 
they are most likely lost from the accelerator. 
The dotted blue line in \Fref{fig:Set-1-fig07} shows probability of photo-ionization loss summed over all possible channels. 
We see the loss rate amounts to $6.9 \times 10^{-10}$, which is found to be dominated by $\sigma_{es}^{(pi)}$, 
the photo-ionization process from $\ket{e}$ by the Stokes laser photons.  
Whether this value is tolerable or not depends on an 
employed acceleration scheme.  To get a feeling, we assume that it takes $10^{-4}$ sec for the ions  
to circle around the accelerator ring ({\it i.e.} its circumference is $C_{ring}=30$ km). 
Then the lifetime of ion beam due to this effect is about $1.4 \times 10^5$ sec: 
this seems long enough to prepare a fresh bunch of ions.

%--------------Subsec 3.3--------------------%
\subsection{Dependence on crucial parameters\\} \label{sec:parameter study}
%--------------Subsec 3.3--------------------%
In this section we will study the dependence of the signal and background rate on crucial parameters 
such as laser intensity, detuning or delay time. 
In all studies below, the He-like Kr is used and only one parameter is varied at a time. 

%
%+++++++++++++++++++++++++++++ figures ++++++++++++++++++++++++++++++
\begin{figure}[htb]
\begin{center}
       \includegraphics[clip,bb=0 0 325 220, width=7.5cm]{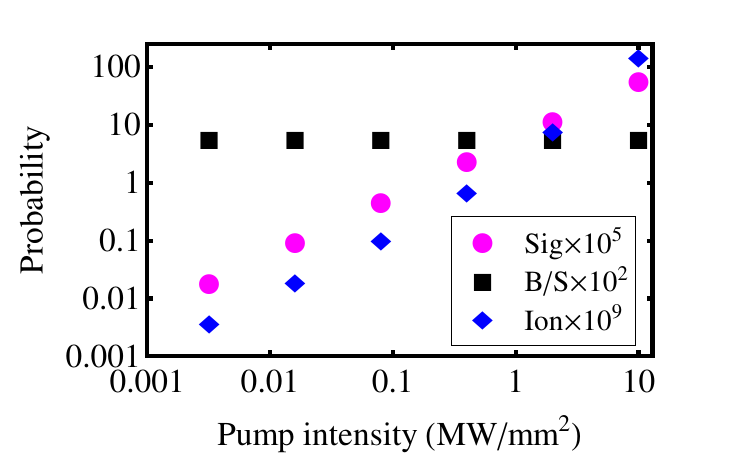}
       \includegraphics[clip,bb=0 0 325 220, width=7.5cm]{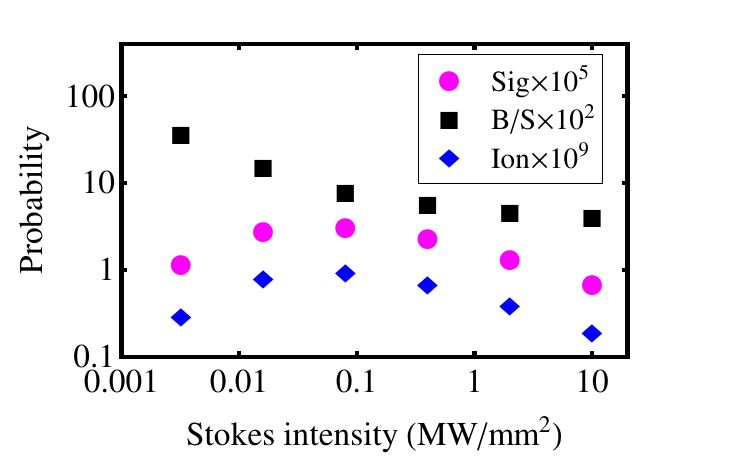}
       \caption{Input power dependence of pump (left) and Stokes (right) lasers. 
       Symbols: signal rate ($\times 10^{5}$, circle in magenta),  
       B/S ratio ($\times 10^{2}$, square in black), 
        ionization rate ($\times 10^{9}$, diamond in blue). 
       }
       \label{fig:pStudy-Kr-Intensity}
\end{center}
\end{figure}
%+++++++++++++++++++++++++++++ figures ++++++++++++++++++++++++++++++
%

\Fref{fig:pStudy-Kr-Intensity} (left) shows the pump laser intensity dependence of the 
signal rate ($\times 10^{5}$, circle in magenta), 
the background-to-signal ratio ($\times 10^{2}$, square in black), 
and the ionization rate ($\times 10^{9}$, diamond in blue).
As seen, in the range we studied, the signal rate increases linearly with the input power while $B/S$
remains constant. The ionization rate also increases with the input power but 
its dependence is faster than linear, and the probability 
reaches $ \sim\! 10^{-7}$ at $I_{p}(0)=10$ MW/mm$^2$ (the right most point), 
leading to a beam lifetime of $\sim\! 1000$ sec. 
\Fref{fig:pStudy-Kr-Intensity} (right) shows a similar plot in case of the Stokes laser intensity.
One notable feature is that B/S decreases monotonically with the laser power, but 
the signal rate peaks around  $I_{s}(0)\simeq 0.1$ MW/mm$^2$. 
The latter feature is caused by an AC Stark effect (light shift): this is confirmed by a two-photon detuning study (see below).

%
%+++++++++++++++++++++++++++++ figures ++++++++++++++++++++++++++++++
\begin{figure}[htb]
\begin{center}
       \includegraphics[clip,bb=0 0 380 220, width=7.5cm]{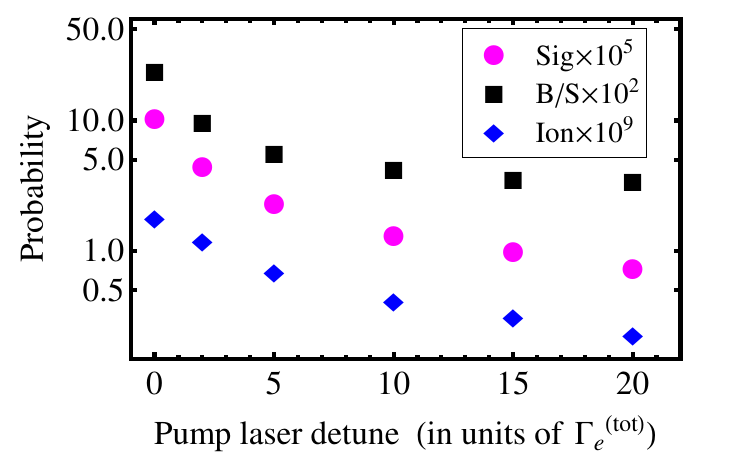}
       \includegraphics[clip,bb=0 0 380 220, width=7.5cm]{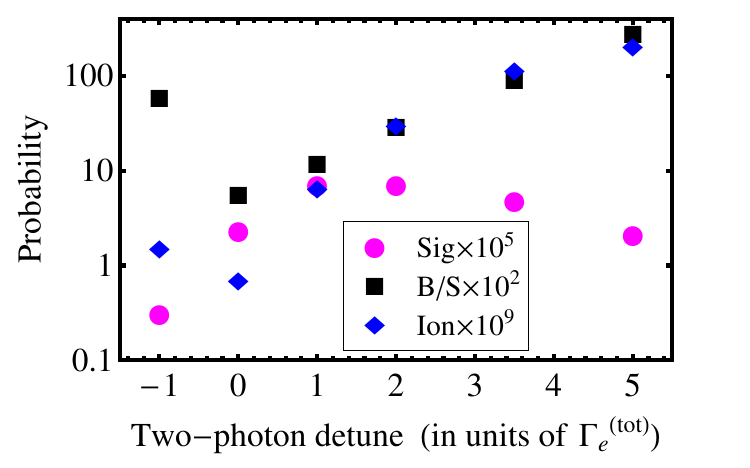}
       \caption{Detuning dependence of pump laser ($\Delta$, left) and two-photon ($\delta$, right). 
       See the caption of \Fref{fig:pStudy-Kr-Intensity} for the symbols.
       }
       \label{fig:pStudy-Kr-detuning}
\end{center}
\end{figure}
%+++++++++++++++++++++++++++++ figures ++++++++++++++++++++++++++++++
%

\Fref{fig:pStudy-Kr-detuning} (left) shows the pump laser detuning ($\Delta$) dependence.
As seen, all quantities decrease monotonically with $\Delta$. 
At $\Delta=0$, the signal increases by a factor of 5, but $B/S$ goes up to $\sim24\%$.
\Fref{fig:pStudy-Kr-detuning} (right) shows a similar plot for two-photon detuning ($\delta$). 
As seen, the signal rate peaks at around $\delta \simeq 1\times \Gamma_{1}^{(tot)}$. 
The phenomenon can be understood as follows. 
It is well known that energy levels may be affected by the AC Stark effect \cite{Cohen-Tannoudji,Foot}. 
In the present case, it is suffice to consider the $\ket{e}$ and $\ket{f}$ two-level system 
since  $\Omega_p$ is much smaller than $\Omega_s$.
The lower state level ($\ket{f}$) would be shifted by $\sim\! \Omega_s^2/(4 \Delta)$: thus 
the actual two-photon resonance condition becomes 
$\delta_{AC}=\omega_p-\omega_s - (\omega_{fg}+\Omega_s^2/(4 \Delta))$, instead of \Eref{eq:def detuning}.
We expect the signal rate is largest when $\delta_{AC}=0$, or $\delta=\Omega_s^2/(4 \Delta)$.
In the present case,  
taking $\Omega_s \sim 60$ ps$^{-1}$ (see \Fref{fig:fig00Rabi}) 
and $\Delta=5 \times \Gamma_{e}^{(tot)}$, we find
$\delta\simeq 1.3 \times \Gamma_{e}^{(tot)}$, 
in rough agreement with the results shown in \Fref{fig:pStudy-Kr-detuning} (right). 
Although the signal rate increases significantly with $\delta$, B/S also increases rapidly.
Thus the choice of $\delta$ must be made considering actual experimental requirements.   

\begin{figure}[htb]
\begin{center}
       \includegraphics[clip,bb=0 0 380 220, width=7.5cm]{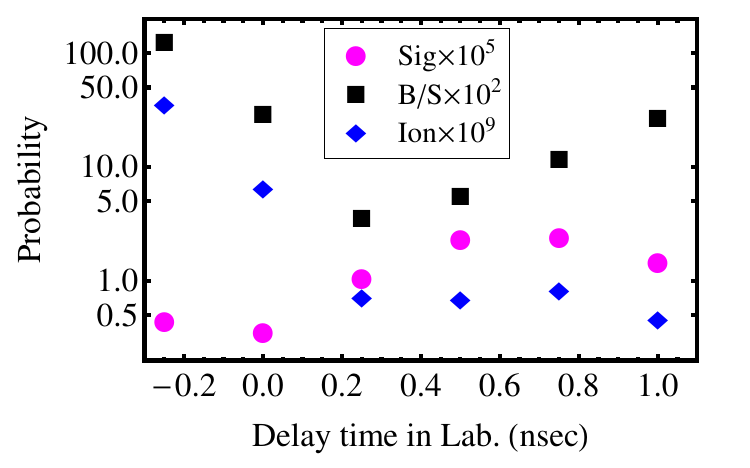}
       %\captionsetup{width=.9\linewidth}
       \caption{Delay time dependence.
       See the caption of \Fref{fig:pStudy-Kr-Intensity} for the symbols.}
       \label{fig:pStudy-Kr-delay}
\end{center}
\end{figure}

Finally the laser delay time dependence is studied as shown in \Fref{fig:pStudy-Kr-delay}.
The signal rate dependence shows a broad peak structure while the $B/S$ ratio has a dip at around $t_d=0.25$ nsec.

%
%
%+++++++++++++++++++++++++++++ figures ++++++++++++++++++++++++++++++
%+++++++++++++++++++++++++++++ figures ++++++++++++++++++++++++++++++
%
%

\subsection{Comparison between Xe and Kr\\}
%--------------Subsec 3.3--------------------%

\begin{figure}[htb]
\begin{center}
       \includegraphics[clip,bb=0 0 380 260, width=7.5cm]{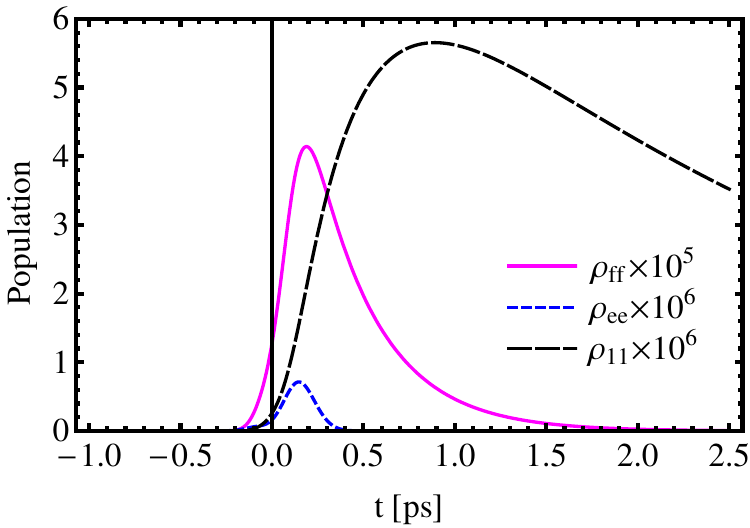}
       \includegraphics[clip,bb=0 0 380 260, width=7.5cm]{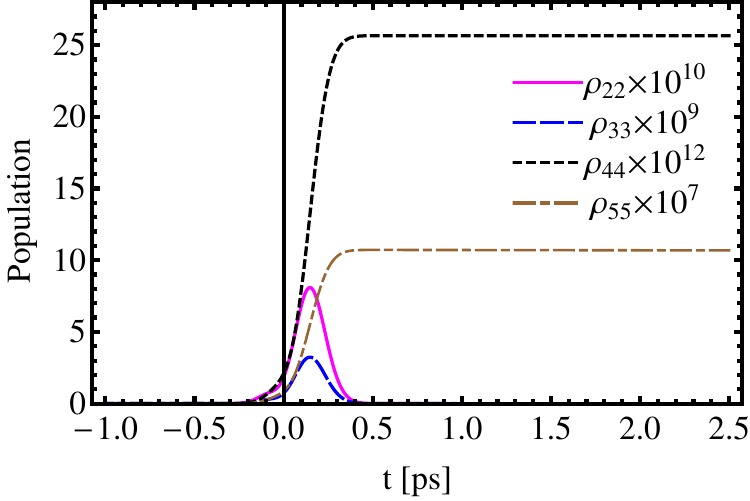}
       %\captionsetup{width=.9\linewidth}
       \caption{Variation of populations as a function of time $t$ in the ion-at-rest frame (Xe). 
       See \Fref{fig:Kr-population} for the symbols, and \Tref{list of parameters sets} (Xe) for the parameters used. 
       }
       \label{fig:Xe-population}
\end{center}
\end{figure}

\begin{figure}[htb]
\begin{center}
       \includegraphics[clip,bb=0 0 380 260, width=7.5cm]{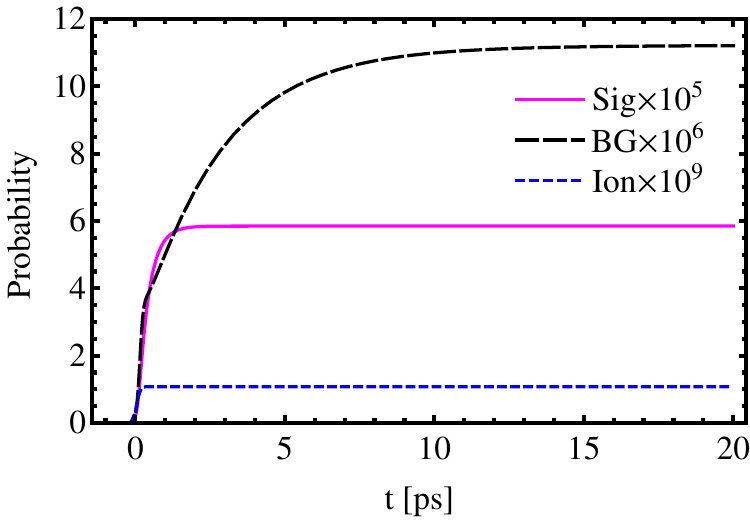}
       %\captionsetup{width=.9\linewidth}
       \caption{Event probability integrated up to time t. 
       See the caption of \Fref{fig:Set-1-fig07} for the symbols, and \Tref{list of parameters sets} (Xe) for the parameters used. 
       }
       \label{fig:Set-2-fig07}
\end{center}
\end{figure}

We now turn our attention to Xe ions. 
\Fref{fig:Xe-population} shows time variation of populations. 
As anticipated, populations decay more quickly than Kr. 
\Fref{fig:Set-2-fig07} shows the probabilities 
of signal, background, and ionization loss integrated up to time $t$. 
In this case, the event rate is defined at $t=20$ psec, considering $\gamma_{ion}$ for Xe 
is twice as large as that for Kr. 
We find the signal and background rates are about $5.8\times 10^{-5}$ and $1.1 \times 10^{-5}$, respectively, 
giving the background-to-signal ratio of 19.2\% while 
the ionization loss rate is about $1.1 \times 10^{-9}$. 
The most significant difference between Xe and Kr is the resulting gamma-ray energies; 
306 MeV for Xe and 65.4 MeV for Kr.
Incidentally, we note that there exists interesting difference between Xe and Kr 
in their  signal and/or background time profile. 
See \Fref{fig:Set-1-fig07} and \Fref{fig:Set-2-fig07}.
In the case of Xe, the signal appears within a few psec after colliding with laser photons while the 
backgrounds grow more slowly. More or less the opposite is true in the case of Kr.
This feature may be utilized in experiments to study differences or characters of the signal and backgrounds. 

%

%+++++++++++++++++++++++++++++ figures ++++++++++++++++++++++++++++++
%+++++++++++++++++++++++++++++ figures ++++++++++++++++++++++++++++++

%
%+++++++++++++++++++++++++++++ figures ++++++++++++++++++++++++++++++
%+++++++++++++++++++++++++++++ figures ++++++++++++++++++++++++++++++
%

%
%--------------Sec 4--------------------%
\section{Discussions and Summary}
%--------------Sec 4--------------------%
In this paper, we presented a new method of generating high-energy gamma rays with orbital angular momentum (OAM).
It utilizes partially-stripped ions (PSIs) as an energy converter: 
accelerated PSIs absorb two photons and emit a photon with OAM. 
When initial PSIs have Lorentz boost factor of $\gamma_{ion}$, 
then the frequency of photons emitted in the backward direction is up shifted in the laboratory to $4\gamma_{ion}^2$ 
times the frequency difference of the two lasers.
One important feature of the proposed method is use of a decay process 
from an excited state with quantum number $J=2,\; M=2$ to a ground state with $J=0,  M=0$. 
Thus emitted photons have an angular momentum $M_\gamma=2$ along the quantization axis (the beam axis).  
An actual scenario of the method is (below the values in the parentheses are those for Xe):
\begin{enumerate}
  \item {Ions:\;} We select He-like Kr (Xe) ions as an energy converter, and populate the 
  	$\ket{2 ^{3\!}P_{2}}$ excited state with the magnetic quantum number $M=2$. 
        Its major decay mode is M2-type radiative transition to the ground state: thus 
        emitted photons have an angular momentum of $M_\gamma=2$.   
  \item {Accelerator:\;} 
  	We assume ions are accelerated up to $\gamma_{ion}=2500$ ($\gamma_{ion}=5000$) 
        in a circular accelerator.  
        A bunched ion beam revolves with a frequency of $10^4$ Hz around its circumference of $C_{ring}=30$ km.  
  \item {Lasers:\;} 
        Two lasers, pump and Stokes, are injected to the ions to realize STIRAP-type population transfer. 
        Their laboratory wavelengths are, respectively, 403 nm and 2690 nm (346 nm and 2371 nm).
        We assume a laser intensity of 400kW/mm$^2$ for each. See \Tref{list of parameters sets} for other parameters.  
\end{enumerate} 
In order to study a signal rate as well as other important features, a set of equations (optical Bloch equations) 
are solved numerically. 
The main findings are as follows. 
\begin{enumerate}
\item {Signal:\;} The signal gamma-ray has an energy of 65.4 MeV (305.9 MeV) in the laboratory, 
	and its rate amounts to $2.3\times 10^{-5}$ ($5.8\times 10^{-5}$). 
	Assuming that there are $10^9$ ions in a bunch, the signal yield is 
	$2.3\times 10^{8}$ Hz ($5.8\times 10^{8}$ Hz). 
	If multiple bunches are put in an accelerator ring, then the yield increases accordingly. 
\item {Background:\;} Background gamma-rays stem from radiative decays from other $n=2$ states, 
	and they have energies of 64.9, 65.6, and 65.1 MeV (301.3, 306.3, and 302.0 MeV),  similar to the signal. 
	The background-to-signal ratio ($B/S$) is found to be 5.6\% (19.2\%). 
	We note that the time profile of the signal and backgrounds differ significantly 
        (see  \Fref{fig:Set-1-fig07} or \Fref{fig:Set-2-fig07}).
	Such difference may be utilized to discriminate the signal from backgrounds, for example, 
	by comparing between early and later parts of gamma-ray pulses.
\item {Ionization loss:\;} Photo-ionization effect might lead to loss of ions in an accelerator ring. 
	In the present case, the loss rate is found to be about $6.9 \times 10^{-10}$ ($1.1 \times 10^{-9}$).
	We conclude that the beam lifetime due to this effect is long enough to prepare a fresh beam.
\item {Parameter dependence:\;} We studied various parameter dependences of the signal rate and $B/S$.
	See section~\ref{sec:parameter study} for details.  
	There are several ways to increase the signal rate further, but usually $B/S$ deteriorates at the same time. 
	Thus parameters must be chosen based on experimental requirements. 
	One exception is the pump laser intensity: in this case 
	the signal rate increases linearly with the input power while keeping $B/S$ constant.
\end{enumerate} 

Finally, we make several comments below.
First, we note that it is possible in principle to choose the $\ket{3 ^{3\!}P_{1}}$ state as $\ket{e}$.
In this case, the states $\ket{g}$ and $\ket{e}$ are connected by the E1 transition while  
$\ket{e}$ and $\ket{f}$ by the E2 transition. 
Unfortunately, there are several disadvantages in this scheme: 
the $B/S$ ratio and ionization rate become much larger in order to obtain a similar level of the signal rate. 
In addition, the Stokes laser requires higher intensity. 

Up to now we have treated rather an idealized ion beam because the main purpose of the 
present paper is to point out the basic principle.
In ref.~\cite{Budker2022}, some representative parameters of the gamma factory are shown. 
One important parameter is the ion beam energy spread ($\Delta \gamma_{ion}/\gamma_{ion}=10^{-4}$), 
which may reduce the signal gamma-ray production rate.
It is expected that the spread in energy would cause the spread in two-photon detuning $\delta$. 
From \Fref{fig:pStudy-Kr-detuning} (right), we find that 
the signal rates in the range $0- 5\times \Gamma_{e}^{(tot)}$ is 
equal to or larger than that of $\delta=0$. 
A very rough estimate of signal reduction factor is given by the effective width of $\delta$ above divided by 
the energy spread of $\ket{f}$, $\hbar \omega_{fg}\times (\Delta \gamma_{ion}/\gamma_{ion})$. 
This turns out to be $\sim\! 3\times 10^{-2}$. 
%$\sim\! 3\times 10^{-2}=\frac{5\cdot(6.68\times 10^{-16})}{(8.40\times10^{-14})\cdot(13090\times 10^{-4})}$. 
Admittedly we need much more detailed analysis or simulation which includes various beam parameters in more detail. 
Another important beam parameters are the transverse beam size and the number of bunches in the ring;  
they are quoted as $16\,\mu$m in radius, and $592-1232$, respectively, in ref.~\cite{Budker2022}. 
Those parameters affect significantly requirements on the lasers, and their 
feasibility must be studied more carefully. 
In any case, such studies are beyond the scope of this paper.

In summary, the proposed method offers an efficient way to produce intense gamma rays with orbital angular momentum.
We hope such a beam would open up new opportunities for a variety of research fields.

\section*{Acknowledgments}
Part of the computation was performed using Research Center for
Computational Science, Okazaki, Japan (Project: 21-IMS-C151).  
MoT was supported by the Inoue Enryo Memorial Grant (Toyo University).
The work of NS is supported in part by JSPS KAKENHI Grant Number
JP 16H02136.
The work of MiT is supported in part by JSPS KAKENHI Grant Numbers 
JP 18K03621,  21H00074, and  21H00168.

\newpage
%--------------Sec 5--------------------%
\section*{Appendix}
%--------------Sec 5--------------------%
%
%\paragraph{Relation between A-coefficients and Rabi frequencies\\}
In this Appendix, we explain how to deduce $\Omega_p$ and $\Omega_s$ from A coefficients. 
We use the Gauss unit system below, which is indicated by the subscript $G$. 

In general, the total E1 (M1) decay rate from $\ket{J_1}$ to $\ket{J_2}$ 
(from $\ket{J_1}$ to $\ket{J_0}$)
 is given by \cite{Messiah}
\begin{eqnarray}
 && A^{(E1)}=\frac{4\, k^3}{3\hbar}\,\frac{1}{2J_1+1}
  \left| \bra{J_1} | \vec{d}_G |\ket{J_2} \right|^2
 \nonumber \\ 
 && A^{(M1)}=\frac{4\, k^3}{3\hbar}\,\frac{1}{2J_1+1}
  \left| \bra{J_1} |\vec{m}_G|\ket{J_0} \right|^2
 \label{eq:E1M1 with reduced matrix element} 
\end{eqnarray}
where $k$ denotes the wavenumber corresponding to the transition, and 
$\bra{J_1} |\vec{d}_G|\ket{J_2}$ ($\bra{J_1} |\vec{m}_G|\ket{J_0}$) is the
reduced matrix element of the electric dipole moment $\vec{d}_G=e_G \vec{r}$ 
(the magnetic dipole moment $\displaystyle \vec{m}_G=\frac{e_G \hbar}{2m_e}(\vec{L}+2\vec{S})$).
The Rabi frequency via E1 (M1) transition, connecting the state 
$\ket{J_1M_1}$ and $\ket{J_2M_2}$ ($\ket{J_0M_0}$ and $\ket{J_1M_1}$), is expressed by 
\begin{eqnarray}
 && \Omega_{R}^{(E1)}=\frac{\bra{J_2M_2}\vec{d_G}\cdot \vec{E}_G \ket{J_1M_1}}{\hbar}
 =\frac{E_{G0}}{\hbar}\bra{J_2M_2} d_G^{\,q} \ket{J_1M_1}
 \nonumber \\
 && \Omega_{R}^{(M1)}=\frac{\bra{J_1M_1}\vec{m}_G\cdot \vec{B} \ket{J_0M_0}}{\hbar}
 =\frac{B_{G0}}{\hbar}\bra{J_1M_1} m_G^{\,q} \ket{J_0M_0}
 \label{eq:E1 and M1 Rabi} 
\end{eqnarray}
where $d_G^{\,q}$ $(m_G^{\,q})$ denotes the $q$-th ($q=\pm1$) component of the dipole operator, 
and $\vec{E}_G$ ($\vec{B}_G$) is the electric (magnetic) field having a polarization $\vec{\epsilon}_q$ 
and magnitude $E_{G0}$ ($B_{G0}$).
Note that $q \;(-q)$ indicates the circular polarization of absorbed (emitted) photon by $\ket{J_1,M_1}$ or $\ket{J_0,M_0}$.
The individual matrix element above and the reduced matrix element are related each other 
by the Wigner-Eckart theorem \cite{Messiah}:
\begin{eqnarray}
 && \bra{J_2M_2} d_G^{\,q} \ket{J_1M_1}
 =\frac{1}{\sqrt{2J_2+1}}\bra{J_2}|d_G|\ket{J_1} C(J_1M_1,1q;J_2M_2),
 \nonumber \\
 && \bra{J_1M_1} m_G^{\,q} \ket{J_0M_0}
 =\frac{1}{\sqrt{2J_1+1}}\bra{J_1}|m_G|\ket{J_2} C(J_0M_0,1q;J_1M_1),
 \label{eq:Wigner-Eckart} 
\end{eqnarray}
where $C(J_1M_1,1q;J_2M_2)$ and $C(J_0M_0,1q;J_1M_1)$ are the Clebsch-Gordan coefficients. 
We now combine \Eref{eq:E1M1 with reduced matrix element}, \eref{eq:E1 and M1 Rabi} and \eref{eq:Wigner-Eckart}, 
and, in the present case, set $q=1$, $J_0=M_0=0$, $J_1=M_1=1$, and $J_2=M_2=2$. 
Inserting $C(00,11;11)=C(11,11;22)=1$, 
we finally arrive at the relations 
\begin{eqnarray}
 && \Omega_{R}^{(E1)}=E_{G0}\sqrt{\frac{9\,A^{(E1)}}{20\, \hbar k^3}}, 
 \hspace{5mm}\mathrm{and}\hspace{5mm}
 \Omega_{R}^{(M1)}=B_{G0}\sqrt{\frac{3\,A^{(M1)}}{4\, \hbar k^3}}.
 \label{eq:Rabi vs A} 
\end{eqnarray}
Note that irrelevant phase factors are omitted above,  
and that the choice of $q=1$ means the right-circular polarization for the pump (absorbed) 
and left-circular polarization for the Stokes (emitted). 
In order to express \Eref{eq:Rabi vs A} by quantities in the SI units, 
we use the conversion rule \cite{Jackson}, $E_{G0}=B_{G0}=\sqrt{4\pi \varepsilon_0}E_0\;$ 
($\varepsilon_0$ being the permittivity of free space), 
along with the fine structure constant 
$\displaystyle \alpha=\frac{e_G^2}{\hbar c}=\frac{e^2}{4\pi \varepsilon_0}\frac{1}{\hbar c}$.

\newpage
\noindent
{\Large References}\vspace{2mm}


\begin{thebibliography}{99}
%
\bibitem{LLQED}
 V.B.~Berestetskii, E.M.~Lifshitz and L.P.~Pitaevskii,
 ``Quantum Electrodynamics'', 2nd Edition, 1980,  Pergamon Press, Oxford. 
 
\bibitem{Allen1992} 
L. Allen, et al. “Orbital angular momentum of light and the transformation of
Laguerre-Gaussian laser modes.”  Phys. Rev. A 45, 8185–8189 (1992).

\bibitem{Shen2019} 
Shen et al. “Optical vortices 30 years on: OAM manipulation from topological charge to multiple singularities”, Light: Science \& Applications 8 :90 (2019)
https://doi.org/10.1038/s41377-019-0194-2

\bibitem{Padgett2017} 
M. J. Padgett, “Orbital angular momentum 25 years on”, Opt. Express 25(10), 11265–11274 (2017).

\bibitem{Molina-Terriza2007} 
G. Molina-Terriza, J. P. Torres and L. Torner, “Twisted photons". Nat. Phys. 3, 305–310 (2007).

\bibitem{Torres-Torner2011} “
Twisted photons”, Edited by J.P.Torres and L.Torner, Wiley-VCH Weinheim, Germany, 2011;  ISBN:978-3-527-40907-5

\bibitem{Babiker2019} Mohamed Babiker, David L Andrews and Vassilis E Lembessis, 
``Atoms in complex twisted light", J. Opt. 21 (2019) 013001


\bibitem{GORGONE2019}
Maria Solyanik-Gorgone, Andrei Afanasev, Carl E. Carlson, Christian T. Schmiegelow, Ferdinand Schmidt-Kaler, 
``Excitation of E1-forbidden atomic transitions with electric, magnetic, or mixed multipolarity in light fields carrying orbital and spin angular momentum",
Journal of the Optical Society of America B, 36, 565 (2019)



\bibitem{Mair2001} A. Mair, A. Vaziri, G. Weihs, and A. Zeilinger, 
``Entanglement of the orbital angular momentum states of photons”, 
Nature 412(6844), 313-316 (2001).

\bibitem{Leach2009} J. Leach, B. Jack, J. Romero, M. Ritsch-Marte, R. W. Boyd, A. K. Jha, S. M. Barnett, S. Franke-Arnold, and M. J. Padgett, 
``Violation of a Bell inequality in two-dimensional orbital angular momentum state-spaces”, Opt. Express 17(10), 8287–8293 (2009).


\bibitem{He1995} H. He, M. E. J. Friese, N. R. Heckenberg, and H. Rubinsztein-Dunlop, 
``Direct observation of transfer of angular momentum to absorptive particles from a laser beam with a phase singularity”, 
Phys. Rev. Lett. 75(5), 826–829 (1995).

\bibitem{Garces2003} V. Garc\'{e}s-Ch\'{a}vez, D. McGloin, M.J. Padgett, W. Dultz, H. Schmitzer, K. Dholakia,  
``Observation of the transfer of the local angular momentum density of a multiringed 
light beam to an optically trapped particle",  
Phys Rev Lett. 2003 Aug 29;91(9):093602. doi: 10.1103/PhysRevLett.91.093602. Epub 2003 Aug 29. PMID: 14525181.

\bibitem{Swartzlander2001}
G. A. Swartzlander,``Peering into darkness with a vortex spatial filter”, Opt. Lett. 26(8), 497–499 (2001).

\bibitem{Swartzlander2008}
G. A. Swartzlander, Jr., E. L. Ford, R. S. Abdul-Malik, L. M. Close, M. A. Peters, D. M. Palacios, and D. W. Wilson,
``Astronomical demonstration of an optical vortex coronagraph”, Opt. Express 16(14), 10200–10207 (2008).

\bibitem{Furhapter2005}
S. F\"{u}rhapter, A. Jesacher, S. Bernet, and M. Ritsch-Marte, 
``Spiral interferometry,” Opt. Lett. 30(15), 1953–1955 (2005).


\bibitem{Erhard2018} Manuel Erhard, Robert Fickler, Mario Krenn and Anton Zeilinger,
``Twisted photons: new quantum perspectives in high dimensions",
Light: Science \& Applications (2018) 7, 17146; doi:10.1038/lsa.2017.146


\bibitem{Gibson2004}
Graham Gibson, Johannes Courtial, Miles J. Padgett, Mikhail Vasnetsov, Valeriy Pas'ko, Stephen M. Barnett, Sonja Franke-Arnold, 
``Free-space information transfer using light beams carrying orbital angular momentum",
Opt. Express 12(22), 5448–5456 (2004)

\bibitem{Krenn2016}
M. Krenn, J. Handsteiner, M. Fink, R. Fickler, R. Ursin, M. Malik, and A. Zeilinger, 
``Twisted light transmission over 143 km”,  
Proc. Natl. Acad. Sci. U.S.A. 113(48), 13648–13653 (2016).



\bibitem{Harwit2003} M. Harwit, ``Photon orbital angular momentum in astrophysics.", 
Astrophys. J. 597, 1266-1270 (2003).

\bibitem{Tamburini2020}
Fabrizio Tamburini , Bo Thide and Massimo Della Valle, 
``Measurement of the spin of the M87 black hole from its observed twisted light",
MNRAS 492, L22–L27 (2020) doi:10.1093/mnrasl/slz176


\bibitem{Maruyama2021}
T.~Maruyama, T.~Hayakawa, T.~Kajino and M.~K.~Cheoun,
``Generation of photon vortex by synchrotron radiation from electrons in 
Landau states under astrophysical magnetic fields'', 
Phys. Lett. B \textbf{826}, 136779 (2022)
doi:10.1016/j.physletb.2021.136779

\bibitem{X-Wang2018} 
Xuewen Wang, Zhongquan Nie, Yao Liang, Jian Wang, Tao Li and Baohua Jia, 
"Recent advances on optical vortex generation", Nanophotonics 2018; 7(9): 1533–1556

\bibitem{Beijersbergen1993} 
M.W. Beijersbergen, L. Allen, H.E.L.O. van der Veen and J.P. Woerdman , 
"Astigmatic laser mode converters and transfer of orbital angular momentum", 
Optics Communications 96 (1993) 123-132

\bibitem{Marrucci2006} 
L. Marrucci, C. Manzo, and D. Paparo, 
“Optical spin-to-orbital angular momentum conversion in inhomogeneous anisotropic media”, 
 Phys. Rev. Lett. 96(16), 163905 (2006)


\bibitem{Sasaki2008} 
S. Sasaki and I. McNulty, “Proposal for generating brilliant x-ray beams carrying orbital angular momentum”,
 Phys. Rev. Lett. 100(12), 124801 (2008).

\bibitem{Bahrdt2013} 
J. Bahrdt, K. Holldack, P. Kuske,  R. M\"{u}ller, M. Scheer,  and P. Schmid, 
``First Observation of Photons Carrying Orbital Angular Momentum in Undulator Radiation",
 Phys. Rev. Lett. 111, 034801 (2013).


\bibitem{Kaneyasu2018} Kaneyasu Tatsuo, Hikosaka Yasumasa,
Fujimoto Masaki, Iwayama Hiroshi, Hosaka Masahito, Shigemasa Eiji, Katoh Masahiro,
``Observation of an optical vortex beam from a helical undulator in the XUV region",
JOURNAL OF SYNCHROTRON RADIATION, 24, 934-938, (2017)

\bibitem{Hemsing2012} 
E. Hemsing, A. Knyazik, F. O'Shea, A. Marinelli, P. Musumeci, O. Williams, S. Tochitsky, and J. B. Rosenzweig, 
``Experimental observation of helical microbunching of a relativistic electron beam",
Appl. Phys. Lett. 100, 091110 (2012); https://doi.org/10.1063/1.3690900

\bibitem{Hemsing2013} 
E. Hemsing, A. Knyazik, M. Dunning, D. Xiang, A.Marinelli, C. Hast, and J. B. Rosenzweig,
``Coherent optical vortices from relativistic electron beams”, Nat. Phys. 9(9), 549 (2013).

\bibitem{Ribic2017} 
P. R. Ribic, B. R\"{o}sner, D.Gauthier, E. Allaria, F.D\"{o}ring, L. Foglia, L. Giannessi, N. Mahne, M. Manfredda, C. Masciovecchio et al., 
``Extreme-ultraviolet vortices from a free-electron laser”, Phys. Rev. X 7(3), 031036 (2017). 

\bibitem{Jentschura2011PRD} 
U. D. Jentschura and V. G. Serbo, 
“Generation of high-energy photons with large orbital angular momentum by compton backscattering”, 
Phys. Rev. Lett. 106(1), 013001 (2011).

\bibitem{Jentschura2011EPC} 
U. D. Jentschura and V. G. Serbo, “Compton upconversion of twisted
photons: Backscattering of particles with non-planar wave functions”, 
Eur.Phys. J. C 71(3), 1571 (2011).

\bibitem{Ivanov-Sebo2011} 
I. Ivanov and V. Serbo, “Scattering of twisted particles: Extension to wave
packets and orbital helicity”, Phys. Rev. A 84(3), 033804 (2011).

\bibitem{Stock2015} 
S. Stock, A. Surzhykov, S. Fritzsche, and D. Seipt, “Compton scattering of
twisted light: Angular distribution and polarization of scattered photons”, 
Phys. Rev. A 92(1), 013401 (2015).

\bibitem{Petrillo2016} 
V. Petrillo, G. Dattoli, I. Drebot, and F. Nguyen, “Compton scattered xgamma
rays with orbital momentum”,  Phys. Rev. Lett. 117(12), 123903
(2016).

\bibitem{Taira2017} 
Y. Taira, T. Hayakawa, and M. Katoh, “Gamma-ray vortices from nonlinear
inverse Thomson scattering of circularly polarized light”, Sci. Rep. 7(1), 5018
(2017).

\bibitem{Chen2019} 
Yue-Yue Chen, Karen Z. Hatsagortsyan, and Christoph H. Keitel, 
"Generation of twisted $\gamma$-ray radiation by nonlinear Thomson scattering of twisted
light", Matter Radiat. Extremes 4, 024401 (2019); https://doi.org/10.1063/1.5086347

\bibitem{Budker2020}
D. Budker {\it et al.}, Ann. Phys. (Berlin) 532 (2020) 2000204.

\bibitem{Tanaka2021}
M. Tanaka and N. sasao, Int. J. Mod. Phys. E 30, (2021) 2150040. 

\bibitem{Serbo2021}
Valeriy G. Serbo, Andrey Surzhykov, and Andrey Volotka, 
Ann. Phys. (Berlin) 534, (2022) 2100199; 
 https://doi.org/10.1002/andp.202100199

\bibitem{Bessonov2013} E.G Bessonov, Nucl. Instr. Meth. B 309, 92 (2013).


\bibitem{Krasny2019}
M.W.~Krasny,
``Gamma Factory, Proof-of-Principle Experiment'',
CERN-SPSC-2019-031; SPSC-I-253.


\bibitem{Ivanov2011} 
P. Ivanov, “Colliding particles carrying nonzero orbital angular momentum”, 
Phys. Rev. D 83(9), 093001 (2011)


\bibitem{Vitanov2017}
N. V. Vitanov {\it et al.}, Rev. Mod. Phys. 89, (2017) 015006.

 
\bibitem{Sobelman}
I. I. Sobelman, ``Atomic Spectra and Radiative Transitions", (Springer, Hederberg 1992). 

\bibitem{Lin1977}
C. D. Lin, R. Johnson, and A. Dalgarno, Phys. Rev. A 15, (1977), 154.


\bibitem{grant}
I.P. Grant,
{\it Relativistic Quantum Theory of Atoms and Molecules},
Springer, New York (2007).

\bibitem{grasp2018}
  C. Froese Fischer, G. Gaigalas, P. J\"onsson and J. Bieron,
  Comput. Phys. Commun. {\bf 237}, 184 (2019).
  
\bibitem{vac pol}
P.J. Mohr, G. Plunien, and G. Soff,
Phys. Rept. {\bf 293}, 227 (1998).


\bibitem{ratip}
 S. Fritzsche,  Comput. Phys. Commun. {\bf 183}, 1525 (2012).


\bibitem{lindblad1976}
  G. Lindblad,  Commun. Math. Phys. {\bf 48} 119 (1976). 


\bibitem{Cohen-Tannoudji}
C.~Cohen-Tannoudji and D.~Guery-Odelin, \textit{Advances in Atomic Physics} 
(World Scientific, Singapore 2011) page-148. 

\bibitem{Foot}
C.~J.~Foot, \textit{Atomic Physics} (Oxford University Press, Oxford 2005) page-144. 

\bibitem{Budker2022}
D. Budker {\it et al.}, Ann. Phys.(Berlin) 534 (2022), 2100284

\bibitem{Messiah}
A.~Messiah, \textit{Quantum Mechanics, Two Volumes Bound as One}  
(Dover, 1999) page-1045, 1052 and 573.

\bibitem{Jackson}
J.~D.~Jackson, \textit{Classical Electrodynamics} 
(Wiley, 1999) page-782
\end{thebibliography}
\end{document}